\documentclass[twocolumn]{aastex631}
\usepackage{silence}
\usepackage{graphicx}
\usepackage[caption=false]{subfig}
\usepackage{booktabs}
\usepackage{stmaryrd}

\usepackage{natbib}
\setcitestyle{authoryear,open={(},close={)}}

\usepackage{float}
\usepackage{color,soul}
\usepackage{wrapfig}
\usepackage{bm}
\usepackage{amsmath}

\usepackage{fancyhdr}
\usepackage{wasysym}

\newcommand{\cii}{[\ion{C}{2}]}

\newcommand{\lir}{L$_{\mathrm{IR}}$}
\newcommand{\lirsf}{L$_{\mathrm{IR,SF}}$}

\newcommand{\loglir}{$\log\mathrm{L}_{\mathrm{IR}}/\mathrm{L}_\odot$}
\newcommand{\sigmaIR}{$\Sigma_{\mathrm{IR}}$}
\newcommand{\sigmaIRSF}{$\Sigma_{\mathrm{IR,SF}}$}
\newcommand{\logSigmaIR}{$\log\Sigma_{\mathrm{IR}}/\mathrm{[L_\odot\,kpc^{-2}]}$}
\newcommand{\logSigmaIRSF}{$\log\Sigma_{\mathrm{IR,SF}}/\mathrm{[L_\odot\,kpc^{-2}]}$}
\newcommand{\sigmaSFR}{$\Sigma_{\mathrm{SFR}}$}

\newcommand{\lpah}{L$_{\mathrm{\tiny PAH}}$}

\newcommand{\lsix}{L$_{6.2\mu\mathrm{\tiny m}}$}

\newcommand{\fagn}{$f_{\tiny\mathrm{AGN,MIR}}$}

\newcommand{\Reff}{$R_{\mbox{\footnotesize eff}}$}

\newcommand{\Mstar}{$\mathrm{M_*}$}

\begin{document}

\title{The IR Compactness of dusty galaxies set star-formation and dust properties at $z\sim0-2$}

\author[0000-0002-6149-8178]{Jed McKinney}
\affiliation{Department of Astronomy, University of Massachusetts, Amherst, MA 01003, USA.}
\affiliation{Department of Astronomy, The University of Texas at Austin, 2515
Speedway Blvd Stop C1400, Austin, TX 78712, USA}

\author[0000-0001-8592-2706]{Alexandra Pope} 
\affiliation{Department of Astronomy, University of Massachusetts, Amherst, MA 01003, USA.}

\author[0000-0002-5537-8110]{Allison Kirkpatrick}
\affiliation{Department of Physics \& Astronomy, University of Kansas, Lawrence, KS 66045, USA}

\author[0000-0003-3498-2973]{Lee Armus}
\affiliation{IPAC, California Institute of Technology, 1200 E. California Blvd., Pasadena, CA 91125, USA}

\author[0000-0003-0699-6083]{Tanio D\'{i}az-Santos}
\affiliation{Institute of Astrophysics, Foundation for Research and Technology-Hellas (FORTH), Heraklion, GR-70013, Greece}
\affiliation{School of Sciences, European University Cyprus, Diogenes street, Engomi, 1516 Nicosia, Cyprus}

\author[0000-0002-4085-9165]{Carlos G\'{o}mez-Guijarro}
\affiliation{AIM, CEA, CNRS, Universit\'{e} Paris-Saclay, Universit\'{e} Paris Diderot, Sorbonne Paris Cit\'{e}, 91191 Gif-sur-Yvette, France}

\author[0000-0002-3560-8599]{Maximilien Franco}
\affiliation{Department of Astronomy, The University of Texas at Austin, 2515
Speedway Blvd Stop C1400, Austin, TX 78712, USA}

\author[0000-0002-7631-647X]{David Elbaz}
\affiliation{AIM, CEA, CNRS, Universit\'{e} Paris-Saclay, Universit\'{e} Paris Diderot, Sorbonne Paris Cit\'{e}, 91191 Gif-sur-Yvette, France}

\author[0000-0003-4073-3236]{Christopher C. Hayward}
\affiliation{Center for Computational Astrophysics, Flatiron Institute, 162 Fifth Avenue, New York, NY 10010, USA}

\author[0000-0003-4268-0393]{Hanae Inami}
\affiliation{Hiroshima Astrophysical Science Center, Hiroshima University, 1-3-1 Kagamiyama, Higashi-Hiroshima, Hiroshima 739-8526, Japan}

\author[0000-0003-1151-4659]{Gerg\"{o} Popping}
\affiliation{European Southern Observatory, Karl-Schwarzschild-Str. 2, D-85748, Garching, Germany}

\author[0000-0003-1207-5344]{Mengyuan Xiao}
\affiliation{AIM, CEA, CNRS, Universit\'{e} Paris-Saclay, Universit\'{e} Paris Diderot, Sorbonne Paris Cit\'{e}, 91191 Gif-sur-Yvette, France}
\affiliation{School of Astronomy and Space Science, Nanjing University, Nanjing 210093, PR China}

\begin{abstract}
Surface densities of gas, dust and stars provide a window into the physics of star-formation that, until the advent of high-resolution far-infrared/sub-millimeter observations, has been historically difficult to assess amongst dusty galaxies. To study the link between infrared (IR) surface densities and dust properties, we leverage the Atacama Large Millimetre/Submillimetre Array (ALMA) archive to measure the extent of cold dust emission in 15 $z\sim2$ IR selected galaxies selected on the basis of having available mid-IR spectroscopy from \textit{Spitzer}. We use the mid-IR spectra to constrain the relative balance between dust heating from star-formation and active galactic nuclei (AGN), and to measure emission from Polycylic Aromatic Hydrocarbons (PAHs) -- small dust grains that play a key role in the photoelectric heating of gas. In general, we find that dust-obscured star-formation at high IR surface densities exhibits similar properties at low- and high-redshift, namely: local luminous IR galaxies have comparable PAH luminosity to total dust mass ratios as high-$z$ galaxies, and star-formation at $z\sim0-2$ is more efficient at high IR surface densities despite the fact that our sample of high$-z$ galaxies 
are closer to the main-sequence than local luminous IR galaxies. High star-formation efficiencies are coincident with a decline in the PAH/IR luminosity ratio reminiscent of the deficit observed in far-infrared fine-structure lines. Changes in the gas and dust conditions arising from high star-formation surface densities might help drive the star-formation efficiency up. This could help explain high efficiencies needed to reconcile star-formation and gas volume densities in dusty galaxies at cosmic noon.
\end{abstract}

\section{Introduction}

The sizes of galaxies are a critical axis along which to study star-formation. In general, the optical/near-IR extent of galaxies gets progressively smaller towards higher-redshifts at fixed star-formation rate and stellar mass \citep{Buitrago2008,Conselice2014,vanderWel2014,Shibuya2015,Mowla2019}, and smaller star-forming galaxies tend to support larger star-formation rate surface densities \citep{Lutz2016,Fujimoto2017}. The sizes of galaxies also correlate with the conditions of the interstellar medium \citep{DiazSantos2017,McKinney2020,McKinney2021a,Puglisi2021}, which may drive changes in the underlying mode of star-formation. Indeed, the scaling relationship between star-formation rate surface densities and molecular gas surface densities is sensitive to the physics of stellar mass assembly in galaxies \citep[e.g.,][]{Schmidt1959,Kennicutt1998b}, and departures from canonical surface density scaling laws have been attributed to changes in the star-formation efficiency \citep{Elbaz2018}. Thus, measuring and accounting for galaxy sizes is an important factor when studying star-formation today and at high-redshift. 

Star-formation from $z\sim0.5-4$ around the peak of the star-formation rate density is predominantly obscured by dust \citep{MadauDickinson2014}, and happens within luminous, infrared galaxies with infrared luminosities (\lir) exceeding $10^{11}\,\mathrm{L}_\odot$ \citep{Murphy2011,Zavala2021}. However, the spatial extent of star-formation in such distant systems was historically difficult to measure in single dish surveys. Until the Atacama Large Millimetre/Submillimetre Array (ALMA) introduced capability for high spatial resolution observations at sub-millimeter wavelengths, studying the extent of star-formation in such dust-obscured galaxies was principally limited by a lack of resolution at infrared wavelengths. Deep radio imaging with the Very Large Array (VLA), Plateau de Bure interferometer, and Submillimeter Array were key in revealing the compact sizes of luminous sub-millimeter galaxies (SMGs) detected in single dish survey \citep[e.g.,][]{Chapman2004,Younger2007,Biggs2008,Tacconi2008,Riechers2011,Bussmann2013}. Recently, much progress has been made towards spatially resolving dust-obscured star-formation on $\sim$kpc scales in $z\sim1-4$ luminous infrared galaxies using ALMA, finding characteristically small sizes $<1-2$ kpc \citep[]{Fujimoto2017,GomezGuijarro2022b,Engel2010,Hodge2016,Ikarashi2015,Spilker2016,Kaasinen2020,Pantoni2021,McKinney2020,Rujopakarn2019,Barro2016} which are reproduced by numerical simulations coupled to far-IR radiative transfer codes \citep{Cochrane2019,Popping2022}. Similarly, low-redshift luminous IR galaxies also show small IR sizes $\lesssim1-3$ kpc \citep{Lutz2016,Bellocchi2022}; however, 
these are commonly confined to merger nuclei whereas high$-z$ targets can show kinematic evidence for disks with high star-formation rate surface densities \citep{Hodge2016,CalistroRivera2018,Pantoni2021,Xiao2022}. The ISM conditions found within such high$-z$ dusty star-forming disks seem to resemble those within the cores of local LIRGs \citep{Spilker2016,McKinney2020,McKinney2021a,Rybak2022}.

From high-resolution ALMA observations, a number of scaling relations between the IR size of galaxies and their star-formation and gas properties have emerged. \cite{Fujimoto2017} statistically demonstrated that sizes measured at IR wavelengths correlate with \lir, and that for fixed \lir\ galaxies at high redshift are on-average smaller than those at low redshift. Sizes measured from dust continuum seem to evolve with stellar mass and redshift in a similar manner as optical sizes for late-type galaxies \citep{GomezGuijarro2022a}, and shrink relative to the stellar light as the gas fraction is diminished \citep{GomezGuijarro2022b}. \cite{Elbaz2018} and \cite{Puglisi2021} target IR-luminous \textit{Herschel} sources at $z\sim1-2$, and find that luminous infrared galaxies with high surface densities tend to have higher star-formation efficiency and higher CO excitation relative to more extended infrared sources at the same redshift. At $z\sim0$ \cite{DiazSantos2017} and \cite{Lutz2016} showed how the IR surface density is a critical axis for understanding key far-IR cooling line emission like \cii\ $157.7\,\mu$m. Radiation field intensities and the far-IR line emission they power depart from typical values above a threshold of $\sim5\times10^{10}\,\mathrm{L_\odot\,kpc^{-2}}$, which may also change the underlying heating and cooling physics in $z\sim0$ luminous IR galaxies \citep{McKinney2021a}. Similar far-IR line ratios and ISM conditions are seen in some high-redshift galaxies with ALMA detections of \cii\ \citep{Zanella2018,Rybak2019,McKinney2020}. Fundamentally, the apparent IR size of dust-obscured star-forming galaxies reflects the surface density of dusty star-forming regions \citep{DiazSantos2017}, and is therefore sensitive to the physical mechanisms regulating gas conditions and star-formation rates.

In this work, we measure IR sizes using archival ALMA observations in a sample of $z\sim2$ galaxies with mid-IR \textit{Spitzer Space Telescope} spectra. Comparing to low-redshift dusty galaxies with similar multi-wavelength observations, we study the link between infrared surface densities (\sigmaIR) and the content and conditions of dust between $z\sim0-3$. 
Building on \cite{Kirkpatrick2017} who look at purely star-forming galaxies, we expand our analysis to include galaxies hosting intermediate to strong buried active galactic nuclei (AGN), the incidence of which within actively star-forming galaxies is high at $z\sim2$ \citep{Sajina2012,Kirkpatrick2012,Kirkpatrick2015}. We decompose infrared emission from galaxies into their star-forming and AGN components using mid-IR \textit{Spitzer} spectra \citep{Pope2008,Kirkpatrick2012}, a key step when accounting for star-formation both with and without AGN. In this manner, we are accounting for galaxies both actively growing their stellar populations and supermassive black holes. From the mid-IR spectra we also measure key dust emission features from polycyclic aromatic hydrocarbons (PAHs) which we compare to the total dust mass as measured by ALMA. 

The paper is organized as follows: In Section \ref{sec:sample} we describe our sample and detail the archival ALMA analysis used to measure IR sizes and dust masses. Section \ref{sec:results} outlines our major results, which we discuss in Section \ref{sec:discussion}. Section \ref{sec:conclusions} summarizes our main conclusions. Throughout this work we adopt a $\Lambda$CDM cosmology with $\Omega_m=0.3$, $\Omega_\Lambda=0.7$, and $H_0=70\,\mathrm{km\,s^{-1}\,Mpc^{-1}}$. We assume a Chabrier IMF \citep{Chabrier2003}. 

\section{Data and Analysis \label{sec:sample}}

In this section we describe the ALMA archival matching process for $z\sim1-2.5$ \textit{Spitzer} targets, our source detection methods, key measured properties, and we comment on the final sample statistics. We also describe similar measurements made for $z\sim0$ comparison samples.  

\subsection{ALMA Archival Sample Selection}

Mid-IR spectroscopy of galaxies is key for decomposing the IR spectral energy distribution (SED) into the components powered by AGN vs.~star-formation. Nuclear toroidal dust heated to high temperatures by buried AGN emits strongly in the mid-IR \citep[e.g.,][]{Laurent2000,Sturm2000,Tran2001}, whereas star-forming regions are bright in broad PAH emission features and exhibit relatively shallower mid-IR spectral indices \citep{Allamandola1989,Pope2008,Sajina2007}. \cite{Kirkpatrick2012} decompose the mid-IR spectra for a large sample by fitting power-law and star-forming templates 
to calculate the $\lambda_{rest}\sim5-12\,\mu$m AGN fraction (\fagn), defined as the fraction of emission within the \textit{Spitzer}/IRS bandpass ($\mathrm{L_{MIR}}$) attributed to an obscured AGN such that $f\mathrm{_{MIR,AGN}=L_{MIR,AGN}/L_{MIR}}$ \citep{Pope2008}. Following \cite{Kirkpatrick2015}, we distinguish between three general \fagn\ categories: star-forming dominated galaxies (SFG, \fagn$\,<20\%$), composite galaxies with intermediary balances between SF and AGN (COM, $20\%<$\fagn$\,<80\%$ ), and AGN dominated galaxies (\fagn$\,>80\%$). Mid-IR AGN with \fagn$\,>80\%$ exhibit a warmer SED, but the average dust temperature of the cold component powered by star-formation is remarkably constant at all \fagn\ around $T_d\sim25$ K \citep{Kirkpatrick2015}. To test the dust-obscured star-formation and dust mass in galaxies over a range of buried AGN strength, we do not select on any \fagn\ threshold. Rather, we use \fagn\ to correct total IR luminosities for the relative contribution from AGN and star-formation. Using Eq.~5 of \cite{Kirkpatrick2015} we first convert the mid-IR AGN fraction to a bolometric IR AGN fraction ($f_{\mathrm{AGN,IR}}$). Next, we determine the IR luminosity attributed to star-formation ($\mathrm{L_{IR,SF}}$) using  $\mathrm{L_{IR,SF}}=(1-f_{\mathrm{AGN,IR}})\times\,$\lir.

Given the unique constraint on dust-obscured AGN and star-formation provided by mid-IR spectroscopy, we select our initial sample of galaxies on the basis of existing \textit{Spitzer}/IRS spectra. Specifically, \cite{Kirkpatrick2012,Kirkpatrick2015} present a parent sample of 151 (Ultra) luminous IR galaxies (LIRGs: \loglir$>11$, ULIRGs: \loglir$>12$) at $z\sim2$ with \textit{Spitzer}/IRS spectra. The original ``supersample'' includes galaxies in the Great Observatories Origins Deep Survey North/South (GOODS-N/S) and is representative of \textit{Herschel+Spitzer} colors of $S_{24\,\mu\mathrm{m}}>0.1\,\mathrm{mJy}$ galaxies with $>3\sigma\,\,250\,\mu$m detections \citep{Sajina2012,Kirkpatrick2012}. Our next selection criterion is on sources that can be observed by ALMA due to their location in the sky, which narrows the candidates from 151 (U)LIRGs across the GOODS fields to 81 (U)LIRGs in GOODS-S. We search for ALMA detections for these 81 GOODS-S galaxies. Galaxies from each \fagn\ classification (SFG, COM, AGN) can be found spanning the redshift and \lir\ range of the parent sample \citep{Kirkpatrick2015}.

\subsection{Source Detection\label{sec:methods}}
We use the following methods to search for ALMA counterparts to the 81 \textit{Spitzer} targets in GOODS-S. We search through the ALMA archive, which includes several large surveys, namely ASAGAO \citep{Hatsukade2018} and GOODS-ALMA 2.0 \citep{GomezGuijarro2022a}. ASAGAO contains a smaller subset of the GOODS-S field than GOODS-ALMA 2.0 but has greater sensitivity. We take as many detections out of ASAGAO, then move to GOODS-ALMA 2.0, and then search the archive for sources not detected in either of the large surveys. We homogenize all ALMA images used in this work to the same imaging parameters; namely, we adopt natural weighting and do not $uv$-taper the data.

For a given $20^{\prime\prime}\times20^{\prime\prime}$ ALMA cutout taken from the archive or the aforementioned large survey maps and centered on the \textit{Spitzer} coordinates, we first derive a local RMS 
after masking potential source emission. Next, we find all peaks above $2.5\sigma$ which we use as priors to create a segmentation map using \texttt{photutils.v1.4} \citep{photutils} \texttt{detect\_sources} with a Gaussian smoothing kernel while enforcing a minimum number of 5 connected pixels, typically less than the number of pixels across the beam FWHM and suitable for flagging spatially unresolved and resolved candidates. We then compare the IRAC4 coordinates against each source found in the segmentation map, and we take the closest match within 1 arcsec for further analysis. As a final check, we next overlay the ALMA contours on top of \textit{Spitzer}/IRAC Ch.~4 (IRAC4) and near-IR imaging from either \textit{JWST}/NIRCam F$_{150W}$ (JADES; \citealt{Eisenstein2023,Rieke2023,JADES}) or \textit{HST}/WFC3 F$_{160W}$ (3D-HST; \citealt{Skelton2014,Grogin2011,Koekemoer2011}) to visually confirm the association. When comparing between ALMA and \textit{HST}, we correct for a global astrometry offset in GOODS-S of $0.^{\prime\prime}09$ in RA and $0.^{\prime\prime}26$ in DEC \citep{Elbaz2018,Franco2018}. We consider all targets with $\geq3\sigma$ contours coincident with the IRAC4 coordinates as candidate ALMA detections for our sample.

Our primary goal is to measure IR sizes from the archival ALMA data to constrain the extent of dust emitting regions in our sample. As discussed in \cite{GomezGuijarro2022a} and \cite{Franco2018,Franco2020}, this requires a continuum peak pixel SNR$\,\geq5$. Therefore, of our candidate archival matches to our sample we only consider detections with SNR$_{\rm peak}>5$ in our analysis.

In summary, of the 81 \textit{Spitzer} targets in GOODS-S we find 23 candidate matches detected in both or either of ASAGAO and GOODS-ALMA 2.0. Of these, 10 are of sufficient SNR to measure an IR size. From searching the archive for observations within 1$^{\prime\prime}$ of our \textit{Spitzer} targets we find 10 more observations with ALMA coverage over our sample of which seven correspond to targets not already detected in the ASAGAO and/or GOODS-ALMA 2.0 maps. Of these seven new matches, five have SNR$>5$ sufficient to measure the IR size and come from the following ALMA programs: 2017.1.01347.S (PI:~A.~Pope, see \citealt{McKinney2020}), and 2018.1.00992.S (PI:~C.~Harrison, see \citealt{Lamperti2021}). Our final sample with measurements of the sub-mm/mm flux and IR size consists of 15 galaxies (10 from ASAGAO+GOODS-ALMA 2.0, 5 from targeted programs). We tabulate the general properties of each galaxy in Table \ref{tab:sample_combined}, and the ALMA-derived quantities are listed in Table \ref{tab:sample_combined}. Near-IR image cutouts with ALMA contour overlays are shown in Fig.~\ref{fig:cutouts}.

\begin{figure}[h!]
    \centering
    \includegraphics[width=.47\textwidth]{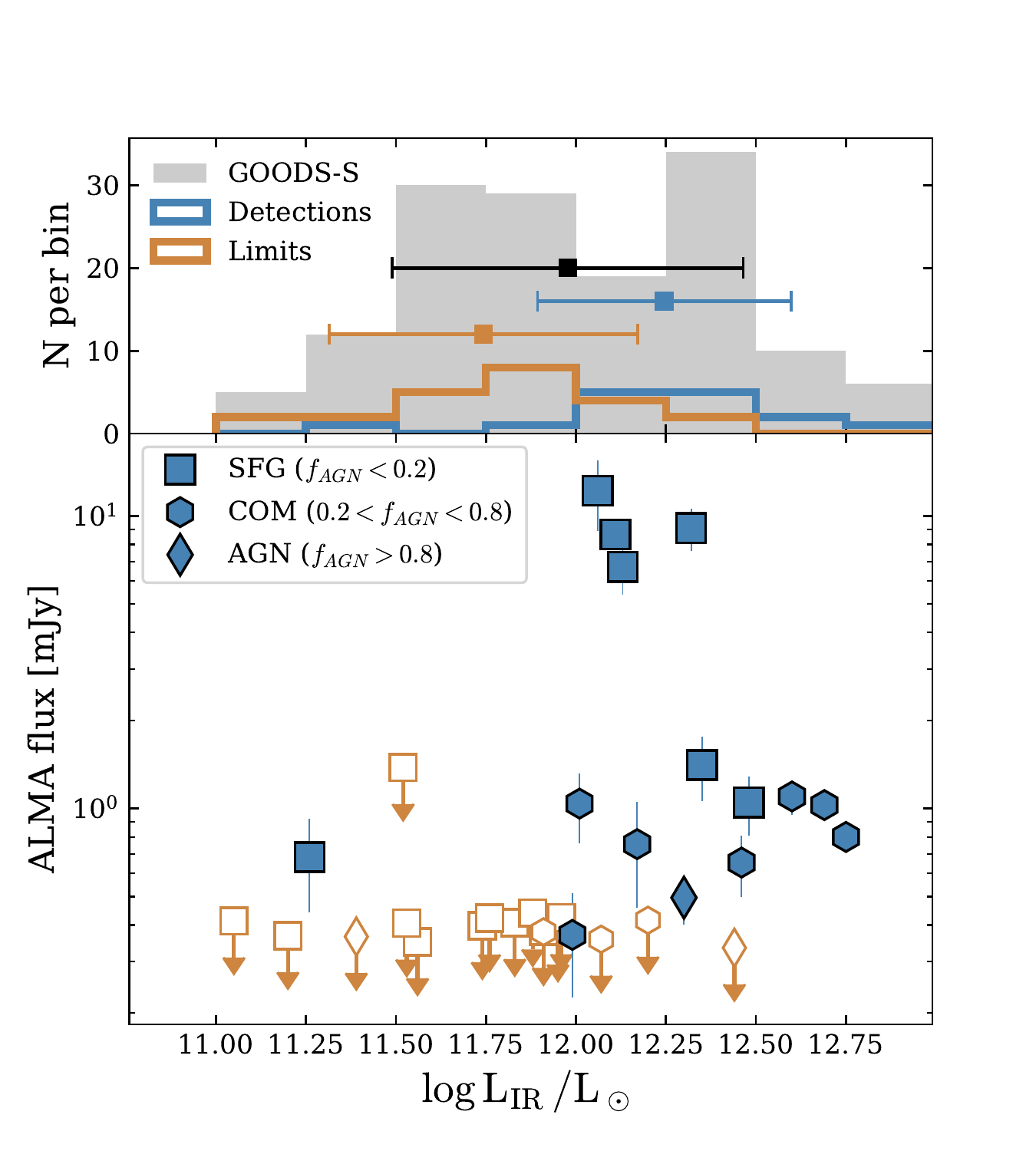}
    \caption{(\textit{Top}) Distribution in \lir\ amongst galaxies matched to ALMA archival detections (blue) compared against non-detections within the footprint of an archival observation (brown) and the parent GOODS-S sample (grey). Colored squares and their errors show the mean and standard-deviation of the corresponding distribution. Our catalog of ALMA-detected sources spans the range in \lir\ of the parent sample, but is biased high on-average by $\sim0.2$ dex. (\textit{Bottom}) Integrated ALMA flux vs. \lir\ for detections and non-detections. Different symbols correspond to the mid-IR AGN classifications as labeled in the caption. Upper limits generally cluster around $\sim0.3$ mJy as most fall within the ASAGAO footprint \citep{Ueda2018,Fujimoto2018}. Note that the four $\sim10$ mJy sources are detected with Band 9 at $\lambda_{obs}\sim450\,\mu$m compared to $\sim1.2\,\mu$m for the rest of the sample. A 25 K blackbody is $\sim10\times$ more luminous at $450\,\mu$m than at 1.2 mm, which is approximately the difference in flux between our Band 9 and Band 6 archival detections. 
    }
    \label{fig:lir_selection}
\end{figure}

\begin{figure}
    \centering
    \includegraphics[width=.47\textwidth]{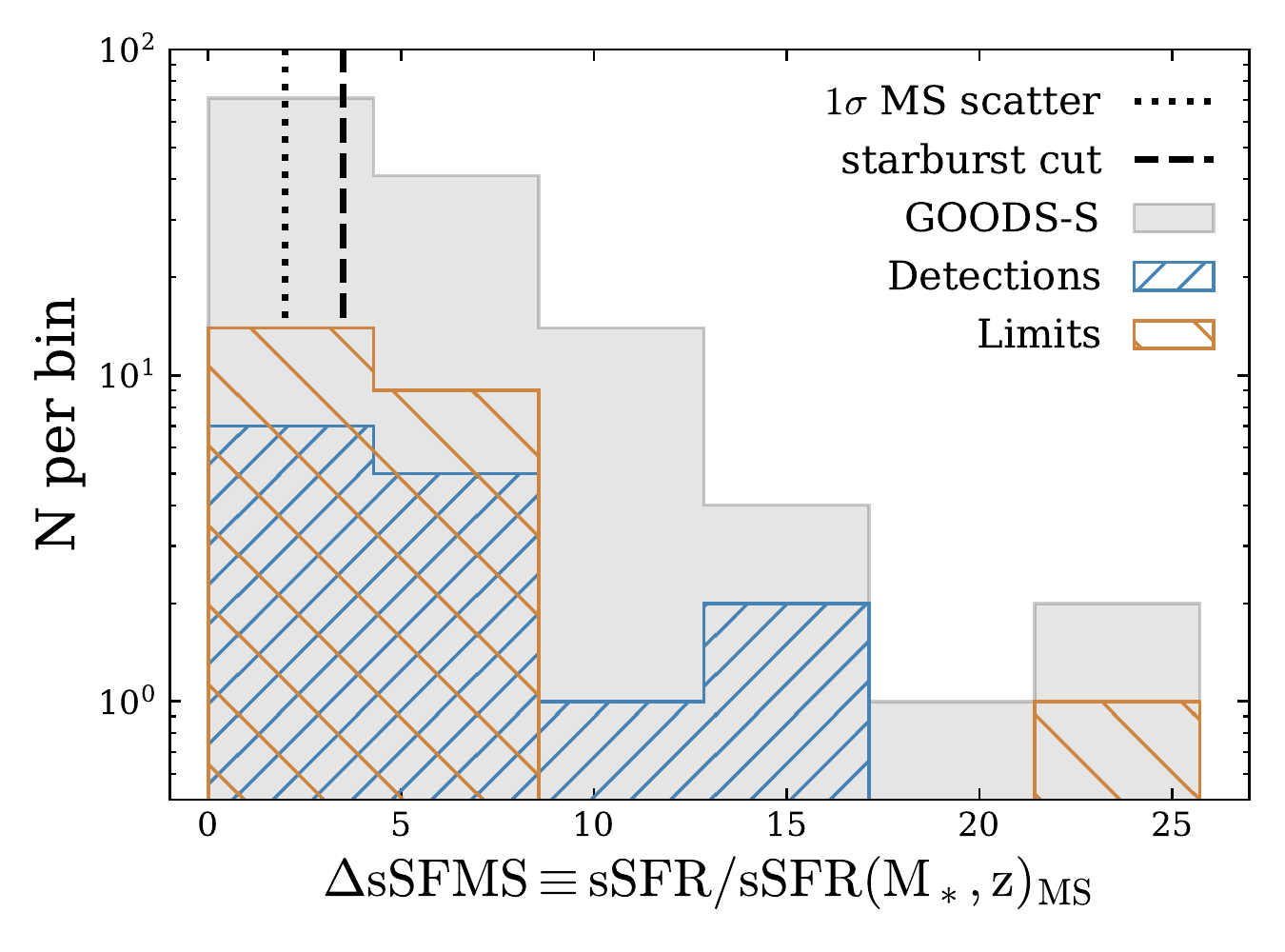}
    \caption{(\textit{Top}) Distribution in specific star-formation rates (sSFR$\,\equiv{\rm SFR/M_*}$) relative to the main-sequence ($\Delta\mathrm{sSFMS}$): the difference between each galaxy's sSFR and the corresponding main-sequence sSFR for its stellar mass and redshift. We adopt the MS parameterization of \cite{Speagle2014}. The color scheme follows the top panel of Fig.~\ref{fig:lir_selection}. ALMA-detected sources have $\Delta\mathrm{sSFMS}$ distributed similarly to undetected archival targets, and consist of 5 galaxies below common thresholds used to identify starbursts (dashed black line, \citealt{Puglisi2021}) and 10 galaxies above.
    }
    \label{fig:sfms_selection}
\end{figure}

Due to the nature of un-targeted archival observations at IR wavelengths, we expect our final ALMA detected sample to be biased towards higher \lir. To test for such bias, we compare the subset of galaxies with robust flux and size measurements from the ALMA archive against sources covered by archival observations but with no detectable signal (Figure \ref{fig:lir_selection}). The mean \lir\ of our final ALMA-detected catalog is $\sim0.2$ dex greater than that of the whole GOODS-S sample, and $\sim0.5$ dex greater than the ALMA non-detections. We detect $84\%$ of all \loglir$\,\geq12$ candidates with coverage in archival observations. While our final catalog is not, on-average, representative of \lir\ and $z$ in GOODS-S 24$\,\mu$m-selected galaxies \citep{Sajina2012,Kirkpatrick2015}, it does span the range of both quantities. As shown in the bottom panel of Fig.~\ref{fig:lir_selection}, we do not preferentially detect any particular mid-IR AGN classification. Most importantly, the ALMA archival detection criterion does not impose a bias on the distribution in specific star-formation rates relative to the main-sequence as shown in Figure \ref{fig:sfms_selection}. Roughly $66\%$ of the galaxies in our final ALMA-detected sample are starbursts (sSFR/sSFR$_\mathrm{MS}>3.5$, \citealt{Puglisi2021}), comparable to the starburst fraction amongst the parent sample and ALMA non-detections. 

\subsubsection{ALMA vs. near-IR morphology and offsets}

Recent data released by the JADES team \citep{Eisenstein2023,Rieke2023,JADES} provides an unprecedented look at the stellar light distribution in dusty galaxies owing to the sensitivity and angular resolution of \textit{JWST}. Nine out of 15 galaxies in our sample are in the JADES/NIRCam map of GOODS-S, and their cutouts are shown on Figure \ref{fig:cutouts}. We find diverse morphologies revealed by \textit{JWST}, ranging from very compact isolated objects (GS IRS1, GS IRS61) to clumpy multi-component distributions likely induced by a merger (GS IRS15, GS IRS20, GS IRS50, GS IRS58, GS IRS60, GS IRS81). The lowest redshift galaxy in our sample GS IRS73 at $z=0.67$ is resolved by NIRCam in exquisite detail, and exhibits spiral arms and a central stellar bulge. The high incidence of irregular morphologies in our sample is consistent with the merger-induced fueling scenario of local luminous, IR-galaxies \citep[e.g.,][]{Hopkins2008}; however, far-IR spectral lines tracing the cold gas kinematics are needed to confirm a merger vs.~clumpy disk scenario.

Offsets between optical/near-infrared and ALMA maps of dusty high$-z$ galaxies can be upwards of a kpc \citep[e.g.,][]{Hodge2015,Chen2017,Simpson2017,CalistroRivera2018,Franco2018}. This could arise from complex dust geometries leading to differential attenuation across the galaxy \citep{Cochrane2021}, and has implications for globally-integrated measures of total star-formation (dust-obscured and obscured) and stellar mass \citep{Simpson2017}. We calculate offsets between rest-frame stellar emission (\textit{JWST, HST}) and cold dust emission (ALMA) across our sample. We measure non-parametric source centroids using the \texttt{photutils} ``center of mass''  centroiding implementation. The galaxy center is therefore defined as the average of pixels over the source weighted by their intensity. Except for GS IRS50 and GS IRS73, every galaxy in our sample has a positional offset between ALMA and \textit{HST} or \textit{JWST} of $<0.^{\prime\prime}3$, consistent with typical ALMA/\textit{HST} offsets reported in the literature \citep[e.g.,][]{Chen2015,Simpson2017,Franco2018}. GS IRS73 is a $z\sim0.7$ spiral galaxy where the dust is preferentially along the northern arm, possibly due to a recent burst of local star-formation. GS IRS50 has a clumpy optical and IR distribution. While it's centroids differ by 5 kpc, peaks in the ALMA map correspond to peaks in the NIRCam image and illustrate a clumpy distribution of both dust-obscured and unobscured star-formation across this galaxy.
Over the whole sample, the 16\%, 50\% and 84\% quantiles on the offsets are 0.6 (0.07), 1.4 (0.17), and 2.5 (0.30) kpc (arcsec) respectively. 

\subsection{Measured Quantities}
\subsubsection{Dust Mass}
Eleven of the archival ALMA observations span a range in wavelength between 870-1250$\,\mu$m, which probes the Raleigh-Jeans tail of cold dust emission over the range in redshifts spanned by our sample ($\sim260-550\,\mu$m between $z\sim1-3$). This regime is aptly suited to measuring the total dust mass because the emission is optically thin at sub-mm wavelengths \citep{Scoville2014}, and the temperature-dependence is linear meaning uncertainties on the mass-weighted dust temperature ($T_{d}$) have modest impact on the total dust mass \citep{Scoville2016,Scoville2017}. Following \cite{Kirkpatrick2017}, we use the ALMA flux densities ($S_\nu$) to measure the dust mass using:
\begin{equation}
    M\mathrm{_{dust}}=\frac{S_\nu D_L^2}{\kappa_\nu B_\nu (T_d)}
    \label{eq:mdust}
\end{equation}
where $D_L$ is the luminosity distance, $B_\nu$ is the Planck equation, and $\kappa_\nu$ is the dust opacity from \cite{WeingartnerDraine2001} assuming MW-like dust and $R_V=3.1$\footnote{At 850$\,\mu$m the dust opacity is $\kappa_{850}=0.15\,\mathrm{m^2\,kg^{-1}}$ \citep{WeingartnerDraine2001}}. As noted by \cite{Kirkpatrick2017}, the variation in $\kappa_\nu$ at longer wavelengths is negligible across common models (e.g.,  MW, SMC, LMC) and for different $R_V$. We choose to fix the cold dust temperature to $T_d=25$ K because most of the dust is cold with a temperature remarkably constant over (1) the full range of mid-IR AGN fractions when the SED is decomposed into its AGN- and SF-powered components using mid-IR spectroscopy \citep{Kirkpatrick2015,Scoville2017}, and (2) redshift for fixed \lir\ \citep{Drew2022}. 

Four of the archival ALMA targets (GS IRS 46, 52, 58, 61) are detected by ALMA at wavelengths below $\lambda_{rest}\sim250\,\mu$m. For two of these targets (GS IRS 46, 52), we use the dust masses derived using 870$\,\mu$m APEX/LABOCA photometry from \cite{Kirkpatrick2017} under the same assumptions and formula as used for longer wavelength detections. Two final sources (GS IRS 58, 61) do not have sub-mm observations along the RJ tail, in which case we place upper limits on the total dust mass using the $3\sigma$ RMS derived from their positions within the ASAGAO map where they are not detected. All dust masses are listed in Table \ref{tab:sample_combined}.

\cite{Kirkpatrick2017} present an analysis of the dust masses of galaxies selected from the \cite{Kirkpatrick2015} supersample on the availability of sub-mm/mm single-dish photometry. Seven of the galaxies we find ALMA archival matches to also have single-dish sub-mm detections in \cite{Kirkpatrick2017}, which we use to test for systematic differences in the dust mass measurements from single-dish and the ALMA interferometer. Flux densities in confusion-limited sub-mm observations are often boosted by the unresolved background as steeply rising source number counts preferentially scatter flux densities upwards \citep[e.g.,][]{Hogg1998,Scott2002,Simpson2015}. Indeed, we find that single-dish derived dust masses tend to be greater than those derived using the ALMA observations by $\sim25\%-50\%$ but both agree within $1\sigma$. 

\subsubsection{Flux Densities and Dust Sizes}
For each target detected in an archival ALMA map, we measure the integrated flux density ($S_{\nu,int}$) and deconvolved (intrinsic) source size by fitting a 2D elliptical Gaussian using \texttt{CASA.imfit}. For four spatially unresolved sources, we use the convolved size (typically negligibly larger than the clean beam) as an upper limit on the extent of the continuum emission, and the peak flux ($S_{\nu,peak}$). For spatially resolved targets, we derive half-light radii (\Reff) from the FWHM by first averaging over the major and minor axes, then using \Reff$\,=\mathrm{\langle FWHM\rangle}/2$. We use \Reff\ to then measure the IR surface density attributed to star-formation ($\mathrm{\Sigma_{IR,SF}}$) using the AGN-corrected total IR luminosities ($\mathrm{L_{IR,SF}}$) and $\mathrm{\Sigma_{IR,SF}=0.5L_{IR,SF}/\pi R_{eff}^2}$.

The size of emission from a source measured by an interferometer can be made in the image-plane after deconvoling the visibilities, or in the $uv-$plane directly on the visibilities. In general, the latter yields a more robust measurement because (1) it avoids uncertainties introduced when reconstructing the sky model during deconvolution, and (2) complex visiblity amplitudes as a function of baseline separation are directly measuring the extent of emission. 
To test consistency between the methods, we measure image-plane sizes for sources in the blindly-selected sample of \cite{GomezGuijarro2022a} using the image-plane method outlined above and compare against their $uv$-plane sizes. The sample of \cite{GomezGuijarro2022a} is taken from GOODS-ALMA2.0 and includes ALMA sources with similar flux densities as our sample at $\sim1.1$mm. When the peak continuum SNR is above $\sim6$, the difference between image- and $uv$-plane sizes is 0.05$^{\prime\prime}$ (400 pc at $z\sim2$) on-average for 18 GOODS-ALMA2.0 sources and both size measures agree within $1\sigma$. This is consistent with previous comparisons in the literature \citep[e.g.,][]{Hodge2016,Chang2020}. Below ${\rm SNR}\sim6$, the difference between image- and $uv$-plane sizes exhibits a larger scatter (0.35$^{\prime\prime}$ for 13 sources) but is 0.07$^{\prime\prime}$ on-average and sizes agree within $1\sigma$ for 80\% of sources at SNR$\,<6$. Sizes are intrinsically uncertain in this lower SNR regime whether made in the image-plane or $uv$-plane \cite[e.g.,][]{GomezGuijarro2022a}. Given the agreement between image- and $uv$-plane sizes for the SNR range spanned by $73\%$ of our sample, we adopt image-plane dust continuum sizes in our analysis with added uncertainty to the lower SNR sub-set.  We add the average uncertainty on $uv$-plane sizes ($0.^{\prime\prime}1$ for SNR$\,<6$) to the image-plane size uncertainties for galaxies in our sample with $5<\mathrm{SNR_{peak}}<6$.

The majority of the ALMA data we use to measure IR sizes were observed at $\lambda_{obs}\sim1.1$ mm (Tab.~\ref{tab:sample_combined}), which at the median $z$ of our sample traces rest-frame $\sim380\,\mu$m emission. One of the high$-z$ galaxies in our sample (GS IRS 20) is spatially resolved at both 450$\,\mu$m and 1.1 mm. 
Following the same procedure outlined in Section \ref{sec:methods}, we measure intrinsic (PSF-corrected) $\lambda_{obs}\sim450\,\mu$m \Reff\ for GS IRS20 to be $\sim20\%$ larger than its 1.1 mm size, and in agreement within $1\sigma$. 
This is consistent with radiative transfer simulations of dusty and massive $z\sim1-3$ galaxies that find a maximal difference of $\sim15\%$ for sizes at $\lambda_{obs}\sim450\,\mu$m vs. $\lambda_{obs}\sim850-1100\,\mu$m \citep{Cochrane2019,Popping2022}. This is close to the accuracy at which we can measure IR sizes. Finally, we choose not to apply any size corrections accounting for variation in the observed wavelengths when comparing our high$-z$ sizes ($\lambda_{\rm rest}\sim350\,\mu$m) against those derived from \textit{Herschel}/PACS in low-$z$ galaxies ($\lambda_{\rm rest}\sim160\,\mu$m). This is physically motivated because the emission is optically thin at $\lambda_{rest}>200\,\mu$m, and the coldest dust component dominating the far-IR emission also dominates the total dust mass \citep{Scoville2017}. 

\begin{deluxetable*}{lccccccccccccccc}
\tabletypesize{\scriptsize}
\setlength{\tabcolsep}{2pt}
\caption{Source characteristics for \textit{Spitzer}/IRS targets matched to archival ALMA observations \label{tab:sample_combined}}
\tablehead{\colhead{ID}&
\colhead{RA}&\colhead{DEC}&
\colhead{$z$\tablenotemark{a}}&
\colhead{$\mathrm{\log\,L_{IR}}$}&
\colhead{$\mathrm{\log\,M_{*}}$}&
\colhead{$f_{AGN}$}&
\colhead{$\mathrm{\log\,L_{6.2\mu m}}$}&
\colhead{$\lambda_{obs}$}&
\colhead{$S_{\nu,int}$}&
\colhead{$S_{\nu,peak}$}&
\colhead{$R_{\mathrm{eff}}$}&
\colhead{$\log M_{\mathrm{dust}}$}&
\colhead{Ref}
\\
\colhead{}&
\colhead{[J2000]}&
\colhead{[J2000]}&
\colhead{}&
\colhead{[$\mathrm{L_\odot}$]}&
\colhead{[$\mathrm{M_\odot}$]}&
\colhead{}&
\colhead{[$\mathrm{L_\odot}$]}&
\colhead{[mm]}&
\colhead{[mJy]}&
\colhead{[mJy/beam]}&
\colhead{[kpc]}&
\colhead{[$\mathrm{M_\odot}$]}&
\colhead{}
}
\startdata
 GS IRS1 & 03:32:44.00 & -27:46:35.0 & 2.69 & 12.69 &  10.95 & 39 &   $9.62\pm0.18$  & 1.233 &  $1.40\pm0.15$ & $0.89\pm0.03$ & $0.62\pm0.16$ & $8.80\pm0.03$ & 1 \\
GS IRS15 & 03:32:40.74 & -27:49:26.0 & 2.11 & 12.17 &  10.78 & 39 &      \nodata     & 1.131 &  $0.76\pm0.30$ & $0.52\pm0.10$ & $1.66\pm0.40$ & $8.44\pm0.17$ & 2 \\
GS IRS20 & 03:32:47.58 & -27:44:52.0 & 1.91 & 12.60 &  10.77 & 25 &   $ 9.90\pm0.18$ & 1.233 &  $0.89\pm0.11$ & $0.53\pm0.05$ & $0.62\pm0.19$ & $8.61\pm0.06$ & 1 \\
GS IRS23 & 03:32:17.23 & -27:50:37.0 & 1.96 & 12.35 &  10.99 &  0 &    $9.11\pm0.99$ & 1.131 &  $1.41\pm0.35$ & $1.01\pm0.15$ & $1.78\pm0.25$ & $8.71\pm0.11$ & 2 \\
GS IRS33 & 03:32:23.43 & -27:42:55.0 & 2.14 & 12.30 &  10.75 & 95 &   $9.10\pm0.10$  & 0.872 &  $0.50\pm0.09$ & $0.18\pm0.02$ & $1.45\pm0.34$ & $8.00\pm0.08$ & 3 \\
GS IRS45 & 03:32:17.45 & -27:50:03.0 & 1.62 & 12.48 &  10.39 &  6 &  $10.19\pm0.06$  & 1.131 &  $1.05\pm0.24$ & $0.81\pm0.09$ & $<3.10$       & $8.46\pm0.06$ & 2 \\
GS IRS46 & 03:32:42.71 & -27:39:27.0 & 1.85 & 12.32 &  10.59 &  0 &  $10.30\pm0.08$  & 0.456 &  $9.10\pm1.50$ & $6.64\pm0.61$ & $1.96\pm1.02$ & $8.70\pm0.16$ & 4\tablenotemark{b} \\
GS IRS50 & 03:32:31.52 & -27:48:53.0 & 1.90 & 12.01 &  10.82 & 28 &   $9.84\pm0.19$  & 1.233 &  $0.07\pm0.03$ & $1.00\pm0.06$ & $<0.64$       & $8.32\pm0.12$ & 1 \\
GS IRS52 & 03:32:12.52 & -27:43:06.0 & 1.79 & 12.11 &  10.43 & 15 &   $9.62\pm0.26$  & 0.444 &  $8.65\pm0.96$ & $5.21\pm0.49$ & $1.53\pm0.85$ & $8.50\pm0.25$ & 4\tablenotemark{b} \\
GS IRS58 & 03:32:40.24 & -27:49:49.0 & 1.85 & 12.06 &  10.86 &  7 &   $9.51\pm0.31$  & 0.456 & 1$2.20\pm3.30$ & $4.86\pm0.93$ & $3.25\pm1.53$ & $<8.35$ & 4\tablenotemark{b} \\
GS IRS60 & 03:32:40.05 & -27:47:55.0 & 2.02 & 12.46 &  10.88 & 23 &  $10.12\pm0.16$  & 1.233 &  $0.65\pm0.16$ & $0.14\pm0.02$ & $3.64\pm0.28$ & $8.48\pm0.10$ & 1 \\
GS IRS61 & 03:32:43.45 & -27:49:01.0 & 1.77 & 12.13 &  10.69 & 15 &  $10.06\pm0.06$  & 0.441 &  $6.70\pm1.30$ & $3.87\pm0.54$ & $1.11\pm0.85$ & $<7.95$ & 4\tablenotemark{b} \\
GS IRS70 & 03:32:27.71 & -27:50:40.6 & 1.10 & 11.99 &  10.78 & 23 & \nodata          & 1.131 &  $0.37\pm0.14$ & $0.35\pm0.06$ & $<2.63$       & $8.05\pm0.10$ & 2 \\
GS IRS73 & 03:32:43.24 & -27:47:56.2 & 0.67 & 11.26 &  10.47 &  0 &   $8.66\pm0.09$  & 1.131 &  $0.68\pm0.24$ & $0.58\pm0.10$ & $1.30\pm0.31$ & $8.20\pm0.15$ & 2 \\
GS IRS81 & 03:32:38.49 & -27:46:31.9 & 2.55 & 12.75 &  10.34 & 38 &   $9.90\pm0.40$  & 1.233 &  $0.60\pm0.12$ & $0.38\pm0.02$ & $0.58\pm0.33$ & $8.43\pm0.03$ & 1 \\
\enddata
\tablecomments{Columns: ($z$) Spectroscopic redshifts derived from fits to the broad PAH features detected in mid-IR \textit{Spitzer}/IRS spectra following Appendix A of \citealt{McKinney2020}. The typical uncertainty on IRS-derived redshifts is $\Delta z\sim0.02$. (\lir) Total IR luminosities derived by fitting \textit{Spitzer} and \textit{Herschel} photometry from \cite{Kirkpatrick2012}, with systematic uncertainties of $\sim10\%$. (\Mstar) Stellar masses originally calculated from optical/near-IR photometry assuming a Salpeter initial mass function \citep{Kirkpatrick2012}, which we have corrected here to a Chabrier initial mass function. (\fagn)  Mid-IR AGN fractions calculated by fitting a star-forming and power-law (AGN) template to the IRS spectra \citep{Pope2008,Kirkpatrick2012}. (\lsix) The luminosity of the $6.2\,\mu$m PAH feature measured from fits to the IRS spectra following App.~A of \cite{McKinney2020}. ($\lambda_{obs}$) Observed continuum effective wavelength. ($S_{\nu,int}$) Source-integrated ALMA flux. ($S_{\nu,peak}$) Peak continuum ALMA flux. ($R_{\mathrm{eff}}$) Effective radius containing half of the total integrated flux. ($M_{\mathrm{dust}}$) Dust mass derived using Eq.~\ref{eq:mdust}. (Ref) ALMA program from which properties are derived: $1=$ASAGAO \citep{Ueda2018}, $2=$GOODS-ALMA \citep{GomezGuijarro2022a}, $3=$2018.1.00992.S \citep{Lamperti2021}, $4=$2017.1.03147.S \citep{McKinney2020}} 
\tablenotetext{a}{ Spectroscopic redshifts are derived from fits to the broad PAH features detected in mid-IR \textit{Spitzer}/IRS spectra (following Appendix A of \citealt{McKinney2020}), and have typical uncertainties of $\Delta z\sim0.02$.}
\tablenotetext{b}{For these objects only detected only in ALMA Band 9 ($\lambda_{obs}\sim450\,\mu$m), we use single-dish dust mass estimates from \cite{Kirkpatrick2017} where possible. Otherwise, we place $3\sigma$ upper limits using local noise properties derived from the target's position within the ASAGAO map.}
\end{deluxetable*}

\subsection{Ancillary Data}
The parent sample from which our targets are selected from have robust multi-wavelength photometry and mid-IR spectroscopy from \textit{Spitzer}/IRS. A full description of the IRS observations can be found in \cite{Pope2008} and \cite{Kirkpatrick2012}, and a comprehensive discussion of the ancillary \textit{Herschel} (PACS and SPIRE) and \textit{Spitzer} (IRAC and MIPS) photometry is presented in \cite{Kirkpatrick2015}. We use the stellar masses derived for our sample in \cite{Kirkpatrick2012} who fit 10 optical/near-IR bands between $U-4.5\,\mu$m with a composite stellar population synthesis code assuming an exponentially declining star-formation history \citep{Drory2004,Drory2009}. \cite{Kirkpatrick2012} assume a Salpeter IMF for their stellar masses, which we convert to a Chabrier framework following \cite{Kirkpatrick2017} ($M_*^{\rm Cha}=0.62M_*^{\rm Sal}$; \citealt{Speagle2014}). Total IR luminosites are derived from fits to \textit{Spitzer}/MIPS and \textit{Herschel}/PACS+SPIRE photometry \citep{Kirkpatrick2012}. The appendix of \cite{McKinney2020} provides a detailed description of how PAH luminosities and spectroscopic redshifts are derived for our sample using a custom Markov Chain Monte Carlo fitting routine. The GOODS-S targets are within the coverage of 3D-HST which provides deep WFC3/IR imaging \citep{Brammer2012,Momcheva2016}. Key derived properties from the ancillary data are listed in Table \ref{tab:sample_combined}. In summary, galaxies in our sample have \lir\ in the range of $10^{11.6}-10^{12.8}\,\mathrm{L_\odot}$, stellar masses between $\sim10^{10}-10^{11}\,\mathrm{M_\odot}$, and redshifts from $z\sim0.7-2.7$.  

\subsection{Comparison Samples\label{sec:comparison_samples}}
We compare our data against local galaxies in the Great Observatories All Sky LIRG Survey (GOALS; \citealt{Armus2009}), a 60$\,\mu$m flux-limited sample of local LIRGs with multi-wavelength data comparable to the coverage of our targets including \textit{Spitzer}/IRS mid-IR measurements of PAH emission \citep{Stierwalt2013,Stierwalt2014}, intrinsic IR sizes from \textit{Herschel}/PACS at $\lambda_{obs}=160\,\mu$m \citep{Lutz2016}, and sub-mm photometry \citep{Chu2017} from which we derive dust masses. \cite{U2012} also present dust massed derived from SED fitting; however, we choose to re-calculate the total dust mass under the same assumptions and with the same method as applied to the high-redshift galaxies to avoid introducing systematic offsets \citep{Kirkpatrick2017}. We use 850$\,\mu$m photometry from the James Clerk Maxwell Telescope where possible to measure the dust mass, and $500\,\mu$m \textit{Herschel}/SPIRE 500$\,\mu$m photometry otherwise \citep{Chu2017}. Dust masses derived from both agree within $\sim20\%$ on average. The \textit{Spitzer}/IRS SL slit usually traces the nuclear region in GOALS \citep{Stierwalt2013}. To estimate galaxy-integrated PAH luminosities in GOALS, we scale luminosity measurements of the PAHs made through the slit by the total-to-slit flux IRAC Ch.~4 flux ratio derived in \cite{Stierwalt2014}. 
Aperture corrections do not correlate with distance \citep{Stierwalt2013} or the total dust mass. 
We use mid-IR AGN fractions in GOALS derived from the 6.2$\,\mu$m equivalent width \citep{DiazSantos2017}, as these most closely resembles our method for measuring \fagn\ at $z\sim2$ where the $6.2\,\mu$m PAH anchors the star-forming template during spectral decomposition \citep{Pope2008}.  

To contextualize the measurements of PAHs in the ISM for both GOALS and our high star-formation rate targets with the population of more normal star-forming galaxies, we compare against galaxies from the KINGFISH survey \citep{Kennicutt2011}, a sample of nearby ($D<30$ Mpc) galaxies  spanning a range in star-formation rate between $0.001-7\,\mathrm{M_\odot\,yr^{-1}}$. The sizes of galaxies in KINGFISH have been reported at optical and FUV wavelengths \citep{Dale2007,Kennicutt2011}, but not in the far-IR. Therefore, we download the \textit{Herschel}/PACS160 maps of KINGFISH targets \citep{Dale2012} from the NASA/IPAC Infrared Science Archive (IRSA) \citep{KINGFISH} and perform a simple aperture-based measurement to derive the effective radii containing 50\% of the 160\,$\mu$m flux. KINGFISH was designed to overlap with existing \textit{Spitzer}/IRS spectroscopy from the SINGS program \citep{Kennicutt2003}, which we use to extend our analysis of dust to low \sigmaIR\ using PAH line fluxes presented in \cite{Smith2007}. We scale the PAH luminosities measured through the IRS slit by the ratio of total \lir\ to \lir\ measured through the slit \citep{Smith2007}, an approximate aperture correction assuming the extent of PAHs follows the cold dust continuum \citep[e.g.,][]{Bendo2008,Calapa2014,Gregg2022}. We note that none of the quantities we derive for KINGFISH correlate with distance or the adopted aperture correction. Finally, we measure dust masses in KINGFISH using \textit{Herschel}/SPIRE $500\,\mu$m photometry \citep{Dale2012,Dale2017} under the same assumptions made for the other data sets. 

PAH line fluxes for GOALS and KINGFISH are measured using spectral decomposition methods (e.g., PAHFIT, \citealt{Smith2007}; CAFE, \citealt{Marshall2007}). Owing to the lower SNR of the $z\sim2$ IRS spectra, we measure PAH line fluxes using a spline continuum fit \citep{Sajina2007,Pope2008,McKinney2020}. \cite{Smith2007} demonstrate that the 6.2$\,\mu$m PAH line luminosities measured with these two techniques differ by a factor of $1.6$ owing to where the continuum is drawn. Therefore, we scale $\mathrm{L_{PAH,6.2}}$ in GOALS and KINGFISH down by a factor of $1.6$ to match our spline-derived PAH luminosities at higher redshift. Omitting this scale factor does not change the results of our analysis, as the empirical scatter in $\mathrm{L_{PAH,6.2}}$ within GOALS, KINGFISH, and $z\sim2$ (U)LIRGs is greater than 60\% of the median. 

\section{Results\label{sec:results}}

\begin{figure}
    \centering
    \includegraphics[width=.47\textwidth]{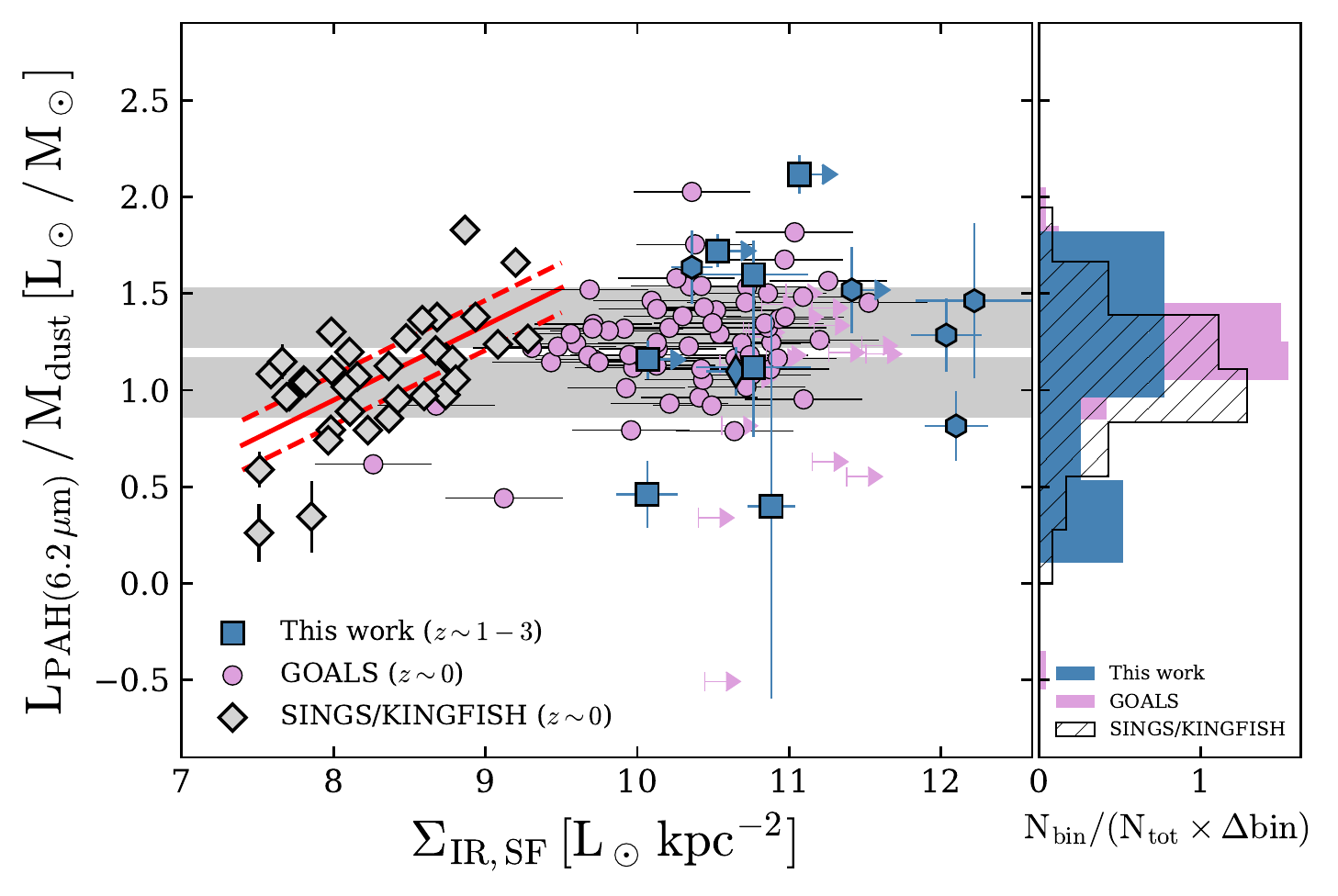}
    \caption{The ratio of the 6.2$\,\mu$m PAH luminosity relative to the total dust mass in our $z\sim2$ ALMA-detected \textit{Spitzer}/IRS sample (blue, following AGN classification symbols of Fig.~\ref{fig:lir_selection}), GOALS (pink circles), and KINGFISH (grey diamonds) as a function of \sigmaIRSF. A histogram showing the distribution in $\mathrm{L_{PAH(6.2\,\mu m)}/M_{dust}}$ for each sample is shown along the right. The scatter in $\mathrm{L_{PAH(6.2\,\mu m)}/M_{dust}}$ is marginally larger for $z>0$ relative to GOALS owing to the lower SNR mid-IR spectra and larger errorbars. Nevertheless, both (U)LIRG samples cluster around an average $\mathrm{\log\,L_{PAH(6.2\,\mu m)}/M_{dust}}\sim1.3\pm0.4\,\mathrm{L_\odot/M_\odot}$ as shown with the shaded region in the left panel with no clear correlation against \sigmaIRSF. KINGFISH galaxies on the other hand do show a positive correlation between $\mathrm{L_{PAH(6.2\,\mu m)}/M_{dust}}$ and \sigmaIRSF\ (solid red line) with small $1\sigma$ dispersion about the best-fit (dashed red line), but the overall range is consistent with ratios found amongst (U)LIRGs. 
    }
    \label{fig:pahs_mdust}
\end{figure}

\subsection{Dust masses}

The dust mass is dominated by large grains which also dominate the far-IR emission, whereas PAHs populate the smaller end of the grain size distribution and emit strong mid-IR features \citep{Draine2001}. Despite this size difference, mid- and far-IR emission tracing the PAHs and cold dust respectively are correlated with the spatial extent of star-formation in galaxies \citep[e.g.,][]{Kirkpatrick2014,Gregg2022} which is, amongst other reasons, why PAHs have been commonly used to trace dust-obscured star-formation rates \citep[e.g.,][]{Genzel1998,Peeters2004,Wu2005,Lutz2007,Pope2008}. In Figure \ref{fig:pahs_mdust} we show the ratio of \lpah\ to total dust mass to empirically trace the PAH mass fraction in the ISM ($q_{\mathrm{PAH}}$, \citealt{Draine2001}) which is otherwise commonly inferred in the literature by fitting dust model grids to spectral energy distributions \citep[e.g.,][]{DraineLi2007,Aniano2020}. We use only the 6.2$\,\mu$m PAH feature because it is isolated from adjacent lines and distant from the strong silicate absorption making it the cleanest PAH line to measure in low SNR \textit{Spitzer}/IRS spectra. We find no correlation between \lpah/$\mathrm{M_{dust}}$ and \sigmaIRSF\ for (U)LIRGs ($p=0.2$), and that all galaxies included in this analysis scatter around an average \lpah-to-total dust mass ratio of $\mathrm{\log\,L_{PAH(6.2\,\mu m)}/M_{dust}}\sim1.2\pm0.3\,\mathrm{L_\odot/M_\odot}$ (Tab.~\ref{tab:par}). This includes galaxies with intermediate to strong dust-obscured AGN, for which spatially resolved \textit{JWST}/MIRI observations have shown to host strong PAH emission remarkably close to the AGN \citep{Lai2022}. We do find a positive correlation between \lpah/$\mathrm{M_{dust}}$ in KINGFISH ($r_p,p=0.63,4.6\times10^{-5}$) with \sigmaIRSF\ which could be driven by the higher metallicities found for warmer, high \sigmaIRSF\ KINGFISH galaxies because of the increasing trend between PAH mass fraction and metallicity \citep{Aniano2020}. The intensity of PAH emission is also a function of metallicity at $z\sim1-2$ \citep{Shivaei2017}, and therefore comparable \lpah/$\mathrm{M_{dust}}$ ratios for GOALS, $z\sim2$ (U)LIRGs and KINGFISH galaxies with \logSigmaIRSF$\,>8.5$ may arise from their similar gas phase metallicities. Given the similar masses of the (U)LIRGs and the mass-metallicity relation, we do not expect any effects of metallicity on the PAH emission for the purposes of comparing GOALS and our $z\sim2$ sample. In any case, the scatter about the best-fit trend in KINGFISH galaxies is substantially lower than for (U)LIRGs at $z\sim0-2$, which could justify the use of PAH emission to trace the total dust mass at \logSigmaIRSF$\,<9.5$ if \sigmaIRSF\ is known. Further studies of high-redshift, low \sigmaIRSF\ galaxies are needed to test if this correlation holds at earlier cosmic epochs. 

\begin{figure*}
    \centering
    \includegraphics[width=\textwidth]{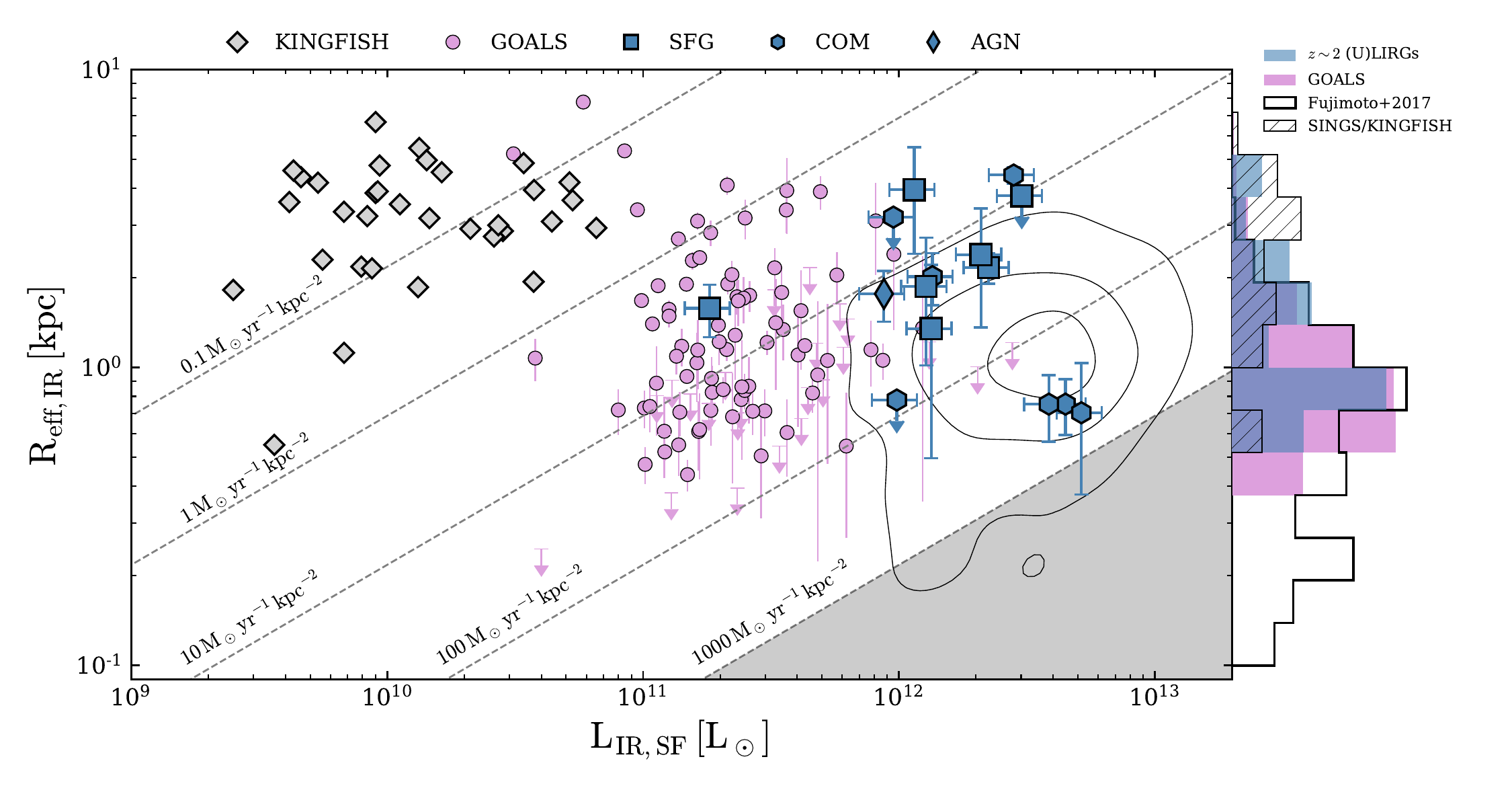}
    \caption{The effective IR size as a function of total infrared luminosity powered by star-formation. Galaxies in our ALMA-detected sample are shown with symbols corresponding to their \fagn\ classification. We compare against the archival sample of \citealt{Fujimoto2017} (black 16\%, 50\%, 84\%\ contours), GOALS (purple circles) and KINGFISH (grey diamonds). We have subtracted out the contribution to \lir\ from AGN in our sample and GOALS using \fagn\ and Eq.~5 from \cite{Kirkpatrick2015}. Dashed lines correspond to constant star-formation rate surface densities. The shaded grey region indicates $\Sigma_{\mathrm{SFR}}>1000\,\mathrm{M_\odot\,yr^{-1}}$ where star-formation exceeds the Eddington limit \citep{Andrews2011}. Normalized histograms of \Reff\ for each sample are plotted along the right $y-$axis. (U)LIRGs have $\Sigma_{\mathrm{SFR}}\sim1-100\,\mathrm{M_\odot\,yr^{-1}}$, with $z\sim2$ galaxies from this work having higher \lirsf\ and greater \Reff\ than the average at $z\sim0$. 
    The three $z\sim2$ composite galaxies close to $\Sigma_{\mathrm{SFR}}\sim1000\,\mathrm{M_\odot\,yr^{-1}}$ are all above the main-sequence for their stellar mass and star-formation rate.
    }
    \label{fig:r_lir}
\end{figure*}

\begin{figure*}
    \centering
    \includegraphics[width=\textwidth]{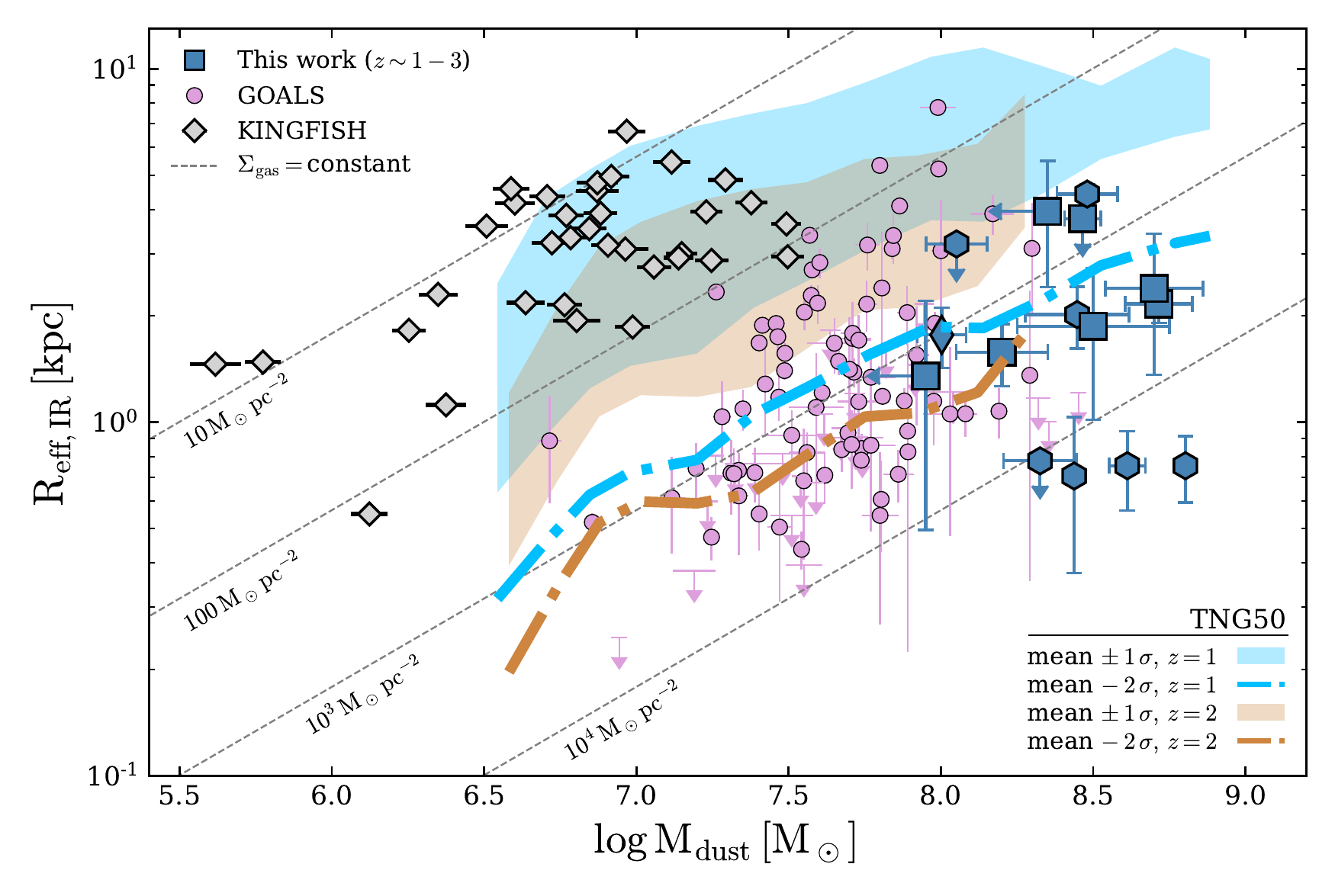}
    \caption{A comparison between the effective IR size and the total dust mass for (U)LIRGs at $z\sim2$ (blue symbols, following \fagn\ classifications in Fig.~\ref{fig:r_lir}), GOALS (pink circles), and KINGFISH (grey diamonds). Grey dashed lines correspond to constant levels of gas mass surface densities  
    assuming a dust-to-gas ratio of 0.01. The shaded regions represent the 16th-84th percentile distribution in dust-continuum size vs. dust mass from galaxies in the TNG50 cosmological simulation at $z\sim1-2$ \citep{Popping2022}. Dot-dashed lines of the corresponding color represent TNG50 galaxies $2\sigma$ below the median. Luminous IR galaxies at $z\sim2$ tend to have higher dust masses and be more extended than their local counterparts in GOALS, except for the three most luminous starbursts in our sample that all have $25\%<$\fagn$\,<40\%$ and have star-formation-rates $>3\times$ greater than main-sequence galaxies for their redshifts and stellar masses. 
    The sizes we measure at $z\sim2$ and those in (U)LIRGs at $z=0$ are consistent with the more compact (lower \Reff) galaxies in TNG50 for fixed dust mass (dashed/dash-dotted lines), and have $\Sigma_{\mathrm{gas}}\sim10^3\,\mathrm{M_\odot\,pc^{-2}}$. 
    }
    \label{fig:r_mdust}
\end{figure*}

\begin{figure}
    \centering
    \includegraphics[width=.485\textwidth]{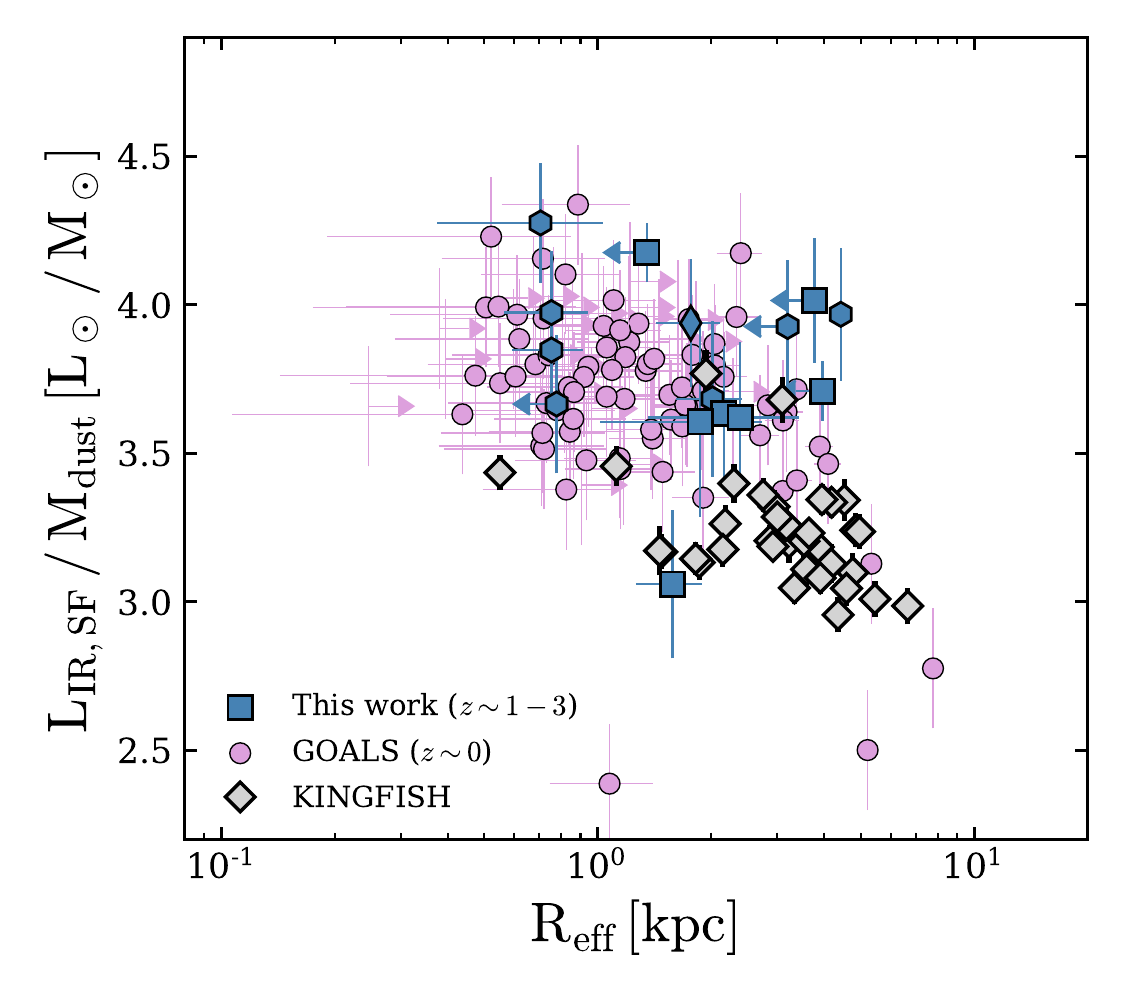}
    \caption{The ratio of total infrared luminosity to total dust mass as a function of IR size in $z\sim0-2$ (U)LIRGs (GOALS, this work) and KINGFISH galaxies at lower star-formation rate surface densities. The legend follows Figures \ref{fig:r_lir} and \ref{fig:r_mdust}. Smaller galaxies in KINGFISH tend to have larger \lirsf/$\mathrm{M_{dust}}$ and eventually fall within the scatter of $z\sim0-2$ (U)LIRGs with low $\mathrm{L_{IR}/M_{dust}}$ by \Reff$\,\sim1-2$ kpc. The anti-correlation between \lirsf/$\mathrm{M_{dust}}$ and \Reff\ at $z\sim0$ including (U)LIRGs and more normal star-forming galaxies is strong $(r_p,p)=(-0.65,10^{-15})$. 
    (U)LIRGs at $z\sim2$ do not exhibit such an anti-correlation, but have comparable $\mathrm{L_{IR}/M_{dust}}$ to local (U)LIRGs albeit at higher \Reff\ on average. 
    } 
    \label{fig:sfe_reff}
\end{figure}

\begin{figure}
    \centering
    \includegraphics[width=.47\textwidth]{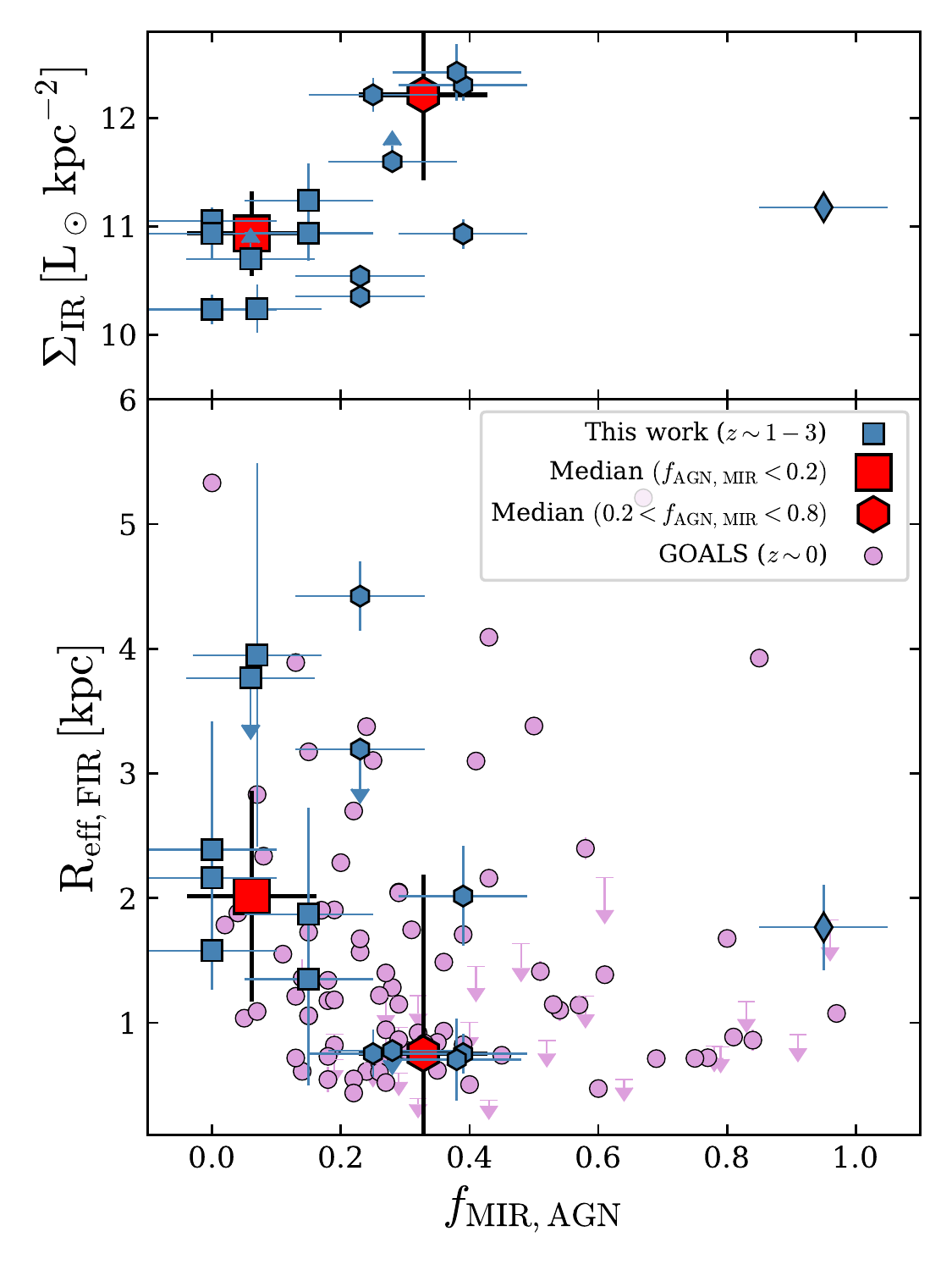}
    \caption{IR surface density not corrected for AGN (\textit{Top}) and IR size (\textit{Bottom}) vs. the fractional contribution of AGN to the integrated mid-IR emission (\fagn). GOALS galaxies are shown with purple circles, and blue squares indicate the $z\sim2$ ALMA-detected sample from this work. 
    At $z\sim2$, dust-obscured galaxies tend to have smaller \Reff\ and higher \sigmaIR\ with increasing \fagn\ on average, exhibiting a factor of $\sim2$ difference in the average sizes between \fagn$\,<0.2$ and  $0.2<$\fagn$\,<0.8$ (red symbols).
    }
    \label{fig:fagn_size}
\end{figure}

\subsection{IR Sizes}
\cite{Fujimoto2017} presents a systematic analysis of IR sizes using Cycles 0-3 ALMA programs made public through the ALMA archive. Their blindly selected sample spans a range in star-formation rates between $\sim10-1000\,\mathrm{M_\odot\,yr^{-1}}$, stellar masses between $\log\,\mathrm{M_*/M_\odot=10-11.5}$ and $z\sim0-6$. We compare the sizes measured in our ALMA-detected sample to their 10$\sigma$-selected sample in Figure \ref{fig:r_lir}, as well as the distribution in sizes found for $z\sim0$ GOALS and KINGFISH galaxies. While we have already subtracted the AGN-powered component from \lir\ using \fagn\ in GOALS and $z\sim2$ (U)LIRGs, this has minimal impact on the \lirsf-\Reff\ relation because $f_{\mathrm{AGN,IR}}$ reaches a maximum of $\sim0.5$ for \fagn$\sim1$ \citep{Kirkpatrick2015}. We have also divided the effective radii reported by \cite{Fujimoto2017} defined as \Reff$\,=0.826\times$FWHM by a factor of $1.6$ for consistency with how we measure \Reff\ to be $0.5\times$FWHM. 
(U)LIRGs from GOALS and our ALMA-detected sample have on-average larger \Reff\ than the blindly selected sample of \cite{Fujimoto2017}, and cluster about a star-formation rate surface density about an order of magnitude smaller. 
Three composite galaxies in our sample are much closer to the locus of \cite{Fujimoto2017} than the rest, and these three also happen to have the highest $\Delta {\rm sSFMS}$ and \lir. This suggests that the offset in sizes between the two samples could be attributed to, in part, observational bias towards brighter objects that preferentially sit above the main-sequence. Indeed, the \cite{Fujimoto2017} sample has a median \lir\ $\sim2\times$ greater than galaxies in our sample.

As shown in Figure \ref{fig:r_lir}, our far-IR size measurements are broadly consistent with the sizes measured for LIRGs and ULIRGs from $z\sim0-6$ spanning a range of star-formation rate surface densities found at all redshifts between $\Sigma_{\mathrm{SFR}}=1-1000\,\mathrm{M_\odot\,yr^{-1}\,kpc^{-2}}$ assuming SFR$_{\mathrm{IR}}/[\mathrm{M_\odot\,yr^{-1}}]=1.49\times10^{-10}\mathrm{L_{IR,SF}/L_\odot}$ \citep{Murphy2011}. (U)LIRGs and ALMA-detected galaxies from this work and \cite{Fujimoto2017} are on average smaller in \Reff\ than KINGFISH galaxies, which tend not to host strong nuclear star-formation as can be found in (U)LIRGs. No galaxy in our $z\sim2$ sample or in GOALS exceeds the theoretical Eddington limit of $\Sigma_{\mathrm{SFR}}\sim1000\,\mathrm{M_\odot\,yr^{-1}\,kpc^{-2}}$ \citep{Andrews2011} even if we omit the AGN correction to \lir. 

In Figure \ref{fig:r_mdust} we show the far-IR extent of the ALMA-detected sample, GOALS, and KINGFISH as a function of their total dust masses. Assuming a dust-to-gas mass ratio of 0.01 as is appropriate for massive $z\sim0-3$ galaxies \citep[e.g.,][]{RemyRuyer2014,Shapley2020,PoppingPeroux2022,Shivaei2022_dust}, we find that most (U)LIRGs would fall along an average $\Sigma_{\mathrm{gas}}\sim1000\,\mathrm{M_\odot\,pc^{-2}}$ at $z\sim0$ and $z\sim2$ as expected from their large star-formation rate surface densities \citep{Kennicutt2012}. The fact that $\mathrm{M_{dust}\propto R_{eff,IR}^2}$ is consistent with optically thin dust \citep{DraineLi2007,Scoville2017}. We note that approximately $25\%$ of GOALS galaxies at $\mathrm{\log\,M_{dust}/M_\odot<8}$ have upper limits on their IR size, which could push the trend towards higher $\Sigma_{\mathrm{gas}}$ at low $\mathrm{M_{dust}}$ amongst the $z\sim0$ LIRG population. 

\cite{Popping2022} derived a simulated distribution in IR size by performing dust radiative transfer on galaxies from the Illustris TNG50 cosmological simulation \citep{Nelson2019,Pillepich2019}. We compare against their simulated IR sizes and dust masses in Figure \ref{fig:r_mdust} which represents galaxies on and above the star-forming main-sequence in TNG50. GOALS galaxies and most of our $z\sim2$ ALMA-detected sample lie above the star-forming main-sequence, and fall within the parameter space in $\mathrm{M_{dust}}$ vs.~\Reff\ spanned by simulated galaxies with sizes $2\sigma$ below the main-sequence trend. Thus, TNG50 reproduces the size and dust mass parameter space observed in (U)LIRGs, but as an outlier population with smaller \Reff\ for fixed $\mathrm{M_{dust}}$ relative to main-sequence galaxies. While the size analysis of \cite{Popping2022} does not extend to $z\sim0$, the positive correlation between \Reff\ and $\mathrm{M_{dust}}$ we find in KINGFISH is comparable to the higher redshift trends for main-sequence galaxies in the simulation (shaded regions on Fig.~ \ref{fig:r_mdust}). 

Figure \ref{fig:sfe_reff} shows the ratio of \lirsf\ to $\mathrm{M_{dust}}$ as a function of the effective IR radius, bringing together the quantities shown independently in Figures \ref{fig:r_lir} and \ref{fig:r_mdust}. The ratio of \lirsf\ to total dust mass is an empirical tracer of the star-formation efficiency, which reflects the amount of star-formation sustained by a galaxy given its total gas content. We find that the \lirsf/$\mathrm{M_{dust}}$ ratios for $z\sim0$ and $z\sim2$ (U)LIRGs are comparable at fixed \Reff, and anti-correlate with the IR size $(r_p,p=-0.65,10^{-15})$. The more spatially extended KINGFISH galaxies have on-average lower \lirsf/$\mathrm{M_{dust}}$ than (U)LIRGs, but reach the lower range of the dustier galaxies below \Reff$\,\sim2$ kpc.

\subsubsection{Far-IR size as a function of mid-IR AGN fraction \label{sec:agn}}

Heavily dust-obscured AGN can produce high IR surface densities as the nuclear torus is heated to high temperatures, emitting predominantly at mid-IR wavelengths. Whether or not colder dust emission emitting at far-IR wavelengths is also powered by AGN remains uncertain and debated in the literature \citep[e.g.,][]{Stanley2017,Scoville2017b,Shangguan2020,Symeonidis2016,Symeonidis2017,Symeonidis2018,McKinney2021b}. Direct heating by AGN and subsequent absorption/re-emission of IR photons could increase the central concentration of IR emission and drive the galaxy-scale effective radii down \citep[e.g.,][]{McKinney2021b,Lamperti2021}. Some star-forming galaxies at high$-z$ do not necessarily exhibit systematically different far-IR sizes compared to sub-mm luminous quasars \citep[e.g.,][]{Chen2021,Ansarinejad2022}; however, prior works do not explicitly control for the AGN strength using mid-IR spectroscopy. 

Using the \textit{Spitzer}/IRS spectral decomposition between SF and AGN for our sample \citep{Pope2008,Kirkpatrick2012} and GOALS \citep{Stierwalt2013,Stierwalt2014,DiazSantos2017}, we show in Figure \ref{fig:fagn_size} the relationship between \sigmaIR\ and \Reff\ with \fagn\ from $z\sim0-2$. GOALS galaxies exhibit no systematic correlation between the far-IR size and \fagn\ consistent with the comparison against PG QSOs in \cite{Lutz2016}. The trend between \fagn\ and \Reff\ could be stronger for the $z\sim2$ ALMA-detected sample in this work. The average \Reff\ decreases by a factor of $\sim2$ and the average \sigmaIR\ increases by $\sim1$dex between \fagn$\,=0-0.4$ for our sample. The highest IR surface density sources are all moderate to strong AGN consistent with \cite{Fujimoto2017} and \cite{Franco2020} who find that X-ray AGN are preferentially found towards greater \sigmaSFR. Further spatially resolved far-IR observations at \fagn$>0.4$ are required to fully test the incidence of small \Reff\ at high \fagn\ in high-redshift galaxies. 
Nevertheless, the most compact and highest IR surface density sources in our sample all have measurable AGN contribution in the mid-IR. 

\begin{deluxetable*}{llccccr}
    \tablecaption{Best-fit Parameters and their Uncertainties for Linear Fits to Data \label{tab:par}}
    \tabletypesize{\footnotesize}
    \tablehead{$x$ & $y$ & $m$ &  $b$ & $\sigma$ & ($r_p,p$) & sample }
    \startdata 
    \logSigmaIRSF & $\mathrm{\log L_{6.2}/M_{dust}\,[L_\odot/M_\odot]}$                           & $0.39\pm0.08$  & $-2.2\pm0.7$  & 0.13  & ($0.63,4.6\times10^{-5}$) & 1    \\
    \logSigmaIRSF & $\mathrm{\log L_{6.2}/M_{dust}\,[L_\odot/M_\odot]}$       &\nodata            & $1.19$         & 0.33          & ($0.13,0.20$)                     & 2,3  \\
    $\mathrm{R_{eff,FIR}\,[kpc]}$  & $\mathrm{\log L_{IR,SF}/M_{dust}\,[L_\odot/M_\odot]}$        & $-0.15\pm0.02$ & $3.87\pm0.04$ & 0.18  & ($-0.65,10^{-15}$)        & 1,2  \\
    \logSigmaIRSF & $\mathrm{\log L_{IR,SF}/M_{dust}\,[L_\odot/M_\odot]}$     & $0.28\pm0.11$     & $0.64\pm1.2$   & 0.14    & ($0.67,0.03$)            & 3     \\
    \logSigmaIRSF & $\mathrm{\log L_{IR,SF}/M_{dust}\,[L_\odot/M_\odot]}$     & $0.25\pm0.02$     & $1.15\pm0.18$  & 0.16    & ($0.79,10^{-26}$)         & 1,2,3 \\
    \logSigmaIRSF & $\mathrm{\log sSFR/sSFR_{MS}}$                            & $0.33\pm0.04$     & $-2.2\pm0.4$   & 0.14     & ($0.67,10^{-14}$)       & 2     \\
    \logSigmaIRSF & $\mathrm{\log sSFR/sSFR_{MS}}$                            & $0.26\pm0.08$     & $-2.25\pm0.9$  & 0.09    & ($0.71,9\times10^{-3}$)  & 3     \\
    \logSigmaIRSF & $\mathrm{\log sSFR/sSFR_{MS}}$                            &\nodata            & 0.64           & 0.27    & \nodata  & 3     \\
    \enddata
    \tablecomments{Fits take the functional form $y=mx+b$. Columns: ($x$) Domain. ($y$) Range. ($m$) Slope. ($b$) $y-$intercept. ($\sigma$) $1\sigma$ dispersion about the best-fit trend. ($r_p,p$) Perason rank coefficient and corresponding $p-$value. (sample) Samples included in fit: (1) KINGFISH, (2) GOALS, (3) ALMA-detected (U)LIRGs at $z\sim2$ from this work. For fits showing no slope the best-fit $b$ and $\sigma$ correspond to the $y-$column average and dispersion about the mean.}
  \end{deluxetable*}

\section{Discussion\label{sec:discussion}}
The dust properties of purely star-forming \textit{Spitzer}/IRS-selected galaxies were studied extensively in \cite{Kirkpatrick2017}. To briefly summarize their main findings, \cite{Kirkpatrick2017} demonstrated that $z\sim2$ galaxies exhibit lower $\mathrm{L_{IR,SF}/M_{dust}}$ ratios and higher gas mass fractions than what are found at $z\sim0$ for fixed \lir, and that galaxies at all redshifts fall along a common  $\mathrm{L_{IR,SF}/M_{dust}}$ relation when accounting for distance above the star-forming main-sequence. We now build upon this prior analysis to understand the role played by the IR sizes of galaxies in governing their star-formation and dust properties. 

\begin{figure}
    \centering
    \includegraphics[width=.5\textwidth]{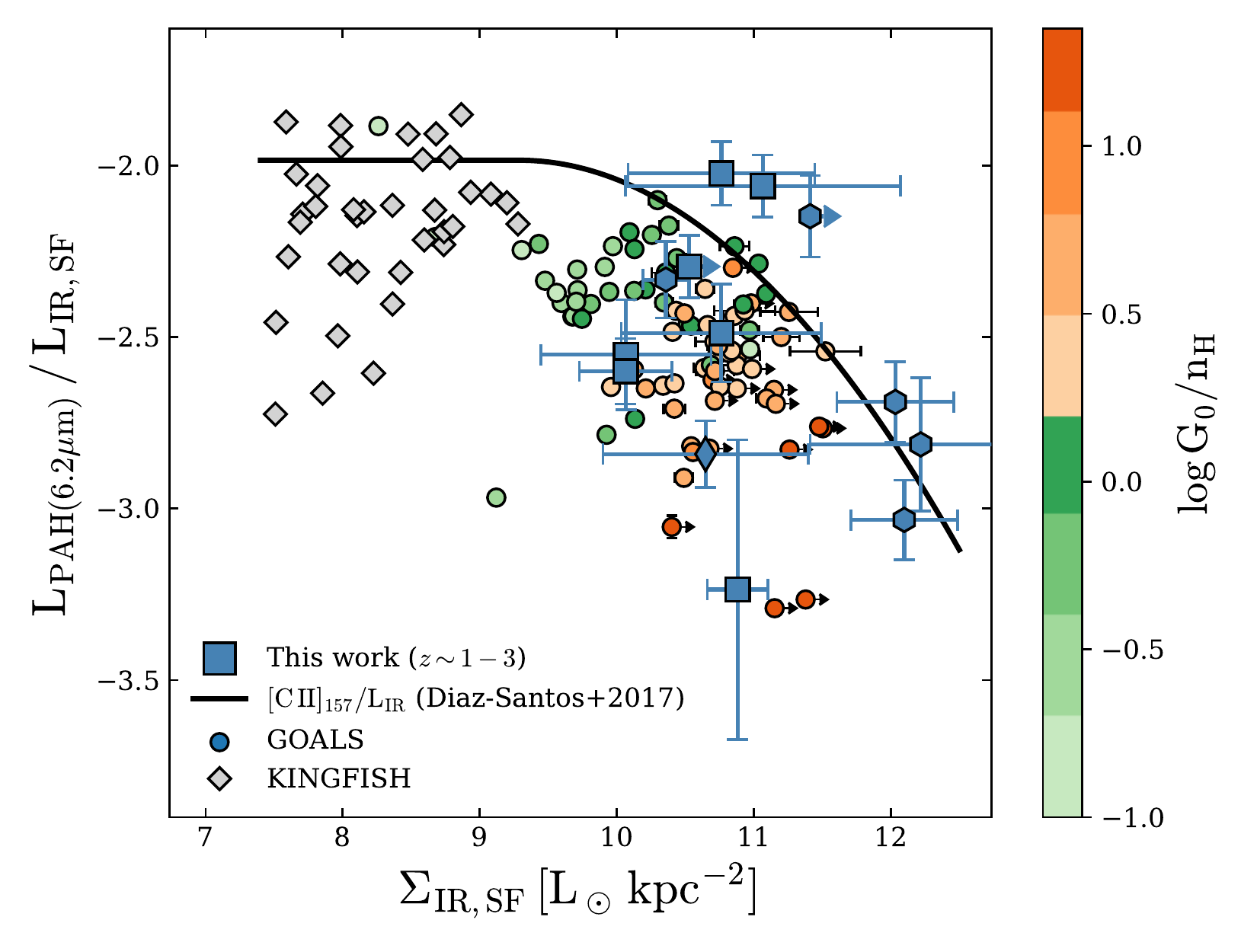}
    \caption{PAH/\lirsf\ vs. IR surface density. GOALS galaxies (circles) are colored by the ratio of the radiation field strength G in units of G$_0$ ( G$_0=1.6\times10^{-3} {\rm erg\, s^{-1}\, cm^{-2}}$, \citealt{Habing1968}) to neutral gas density ($\mathrm{n_H /cm^{-3}}$) in PDRs, derived from FIR find-structure line modeling \citep{DiazSantos2017}. For comparison, we also show the trend in \cii$_{\mathrm{neutral}}$/\lir\ from \cite{DiazSantos2017} as a black line which has not been scaled. KINGFISH (grey diamonds) cluster at lower \sigmaIRSF\ about this trend. (U)LIRGs at $z\sim2$ (blue squares) follow the local trend in PAH/\lirsf\ vs. \sigmaIRSF, which exhibits a comparable turnover as found for the deficit observed in \cii\ emission at \logSigmaIR$\,>10$ accompanied by high G/$\mathrm{n_H}$. This suggests a change in the ISM conditions with \sigmaIRSF\ for $z\sim2$ dusty, star-forming galaxies comparable to the trends found with IR surface density in GOALS galaxies \citep{McKinney2021a}.
    }
    \label{fig:sigma_PAH}
\end{figure}

\begin{figure*}
    \centering
    \gridline{
        \fig{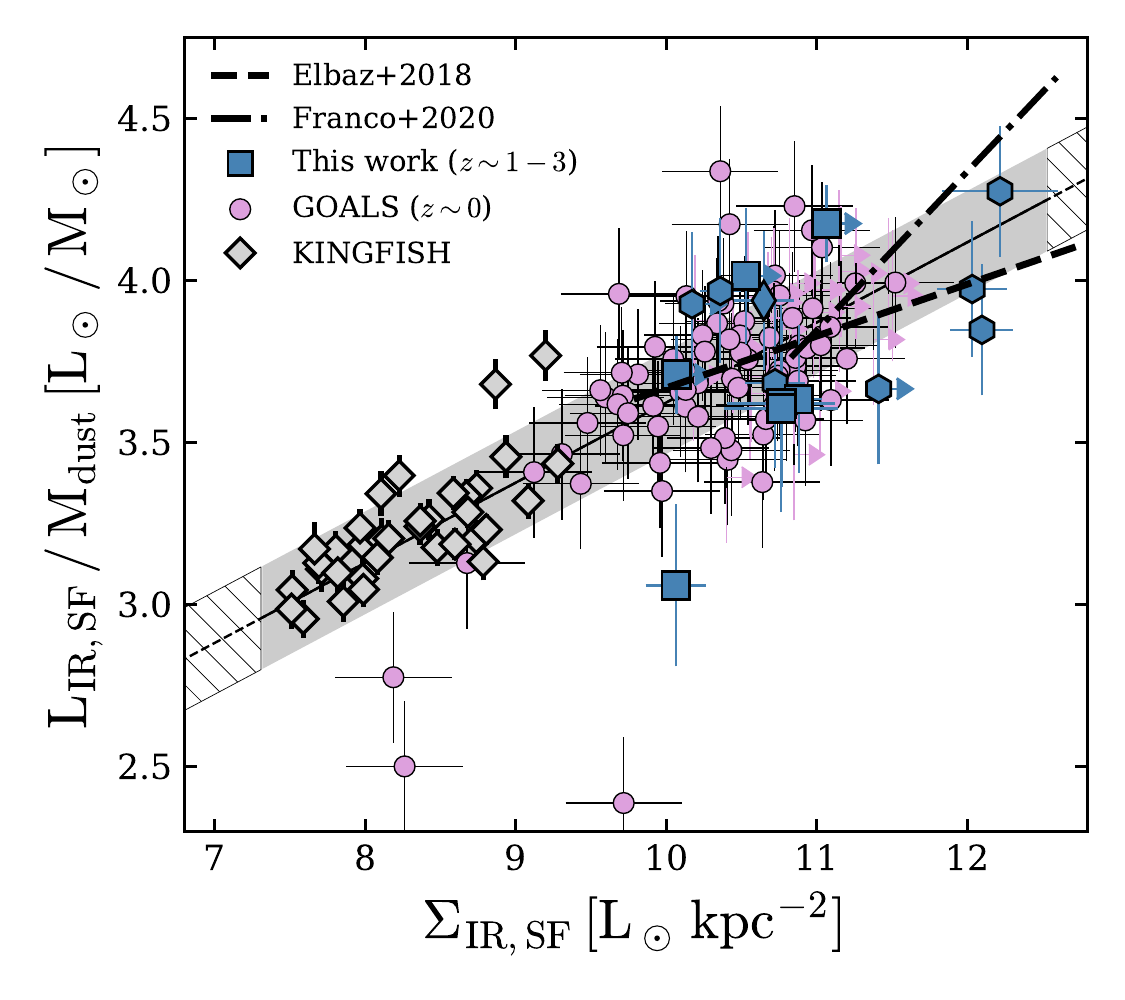}{0.49\textwidth}{}
        \fig{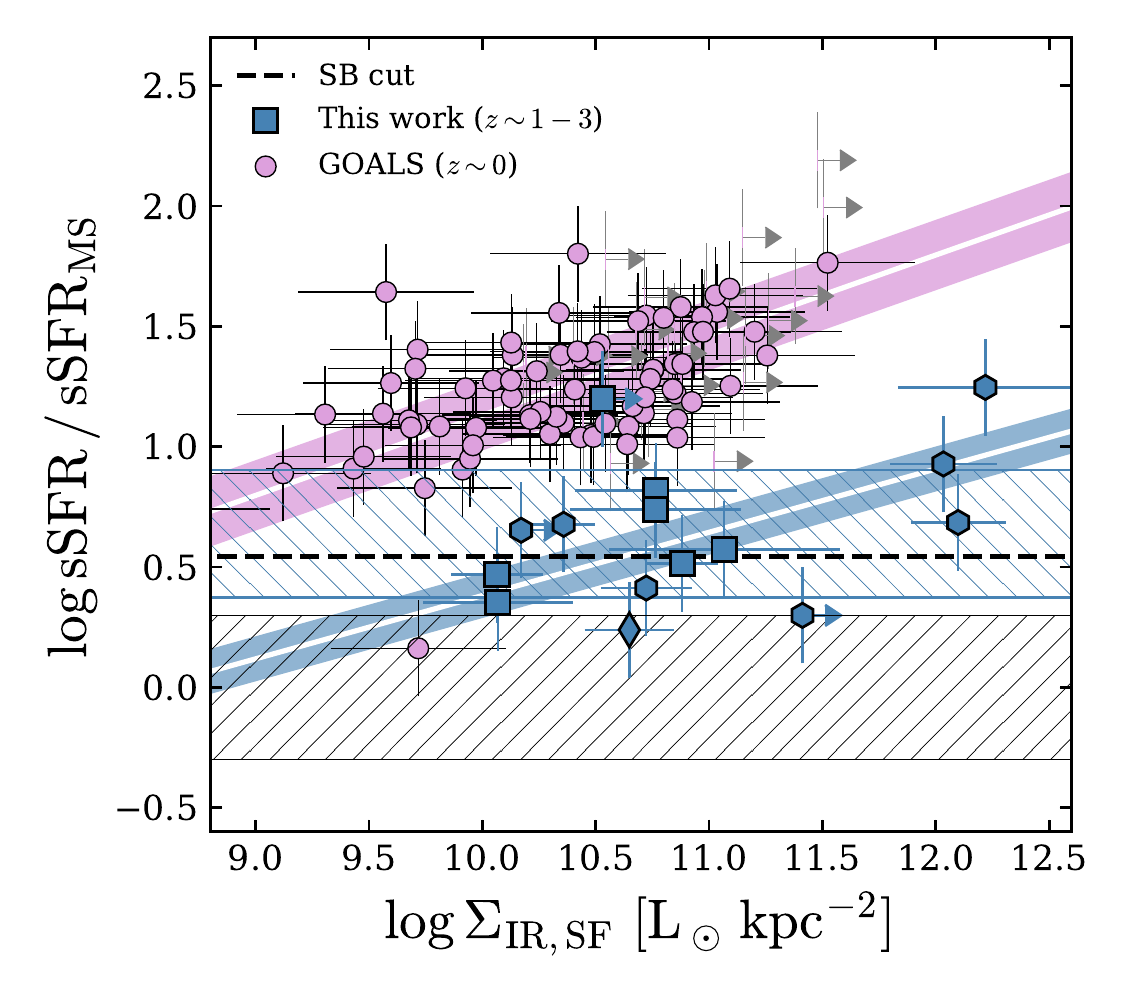}{0.49\textwidth}{}
    }\vspace{-30pt}
    \caption{(\textit{Left}) The ratio of AGN-corrected IR luminosity to total dust mass, a tracer of the star-formation efficiency, as a function of IR surface density. Labeling follows Fig.~\ref{fig:r_lir}. A best-fit linear trend to all galaxies on the Figure is shown in solid grey. Hatched regions show the fit beyond the domain of our data. The dashed and dot-dashed line shows the SFE and IR size relation from \cite{Elbaz2018} derived from $870\,\mu$m imaging of \textit{Herschel}-selected (U)LIRGs, and from mm-selected galaxies in \cite{Franco2020} respectively, each over their respective domains in \sigmaIR. Both literature trends are normalized to the locus of our data as differences in SFR- and gas-mass derivations can change the absolute normalisation of the trend. We find a redshift-independent relation between the IR size and \lirsf/$\mathrm{M_{dust}}$ ratio where higher IR surface density galaxies exhibit evidence for higher star-formation efficiencies. 
    (\textit{Right}) Distance from the specific star-forming main-sequence for each galaxy's redshift and stellar mass as a function of \sigmaIRSF. The black hatched region encases the $\pm0.3$ dex canonical main-sequence scatter \citep{Whitaker2014}, and the dashed black line indicates the threshold above which galaxies are considered starbursts \citep{Puglisi2021}. (U)LIRGs at high IR surface density tend to be further above the main-sequence, a trend that is more pronounced at $z\sim0$ (shaded pink line) than at $z\sim2$ (shaded blue line) where the data is well-fit with no \sigmaIRSF-dependence (hatched blue region).
    }
    \label{fig:discussion}
\end{figure*}

\subsection{ISM conditions and dust composition scale with IR surface densities\label{sec:discussion:ism}}

For fixed \lir, galaxies at $z\sim2$ exhibit more PAH emission per unit \lir\ than local galaxies \citep{Pope2013}. In a small sample of ALMA-selected \textit{Spitzer} targets, \cite{McKinney2020} demonstrated that this offset disappears for fixed \sigmaIR. We expand upon this prior result using a sample three times larger and sizes predominantly measured along the RJ tail of cold dust emission. Figure \ref{fig:sigma_PAH} shows the \lpah/\lirsf\ ratio as a function of \sigmaIRSF\ at $z\sim2$, and for GOALS and KINGFISH. We find the same \lpah/\lirsf\ ratios at high and low redshift for fixed \sigmaIRSF. 
Moreover, we recover an anti-correlation between \lpah/\lirsf\ and \sigmaIRSF\ reminiscent of photometric measures of PAH emission in dusty galaxies (i.e., IR8; \citealt{Elbaz2011}), and the FIR fine-structure line deficit observed in low- and high$-z$ dusty galaxies \citep{DiazSantos2017,Zanella2018,McKinney2020}. PAHs and FIR lines predominantly arise from photodissociation regions (PDRs) around sites of recent star-formation for actively star-forming galaxies \citep{Tielens1985,Malhotra1997,Malhotra2001,Tielens2008,Beirao2012,Croxall2017,DiazSantos2017,Sutter2019}, and thus the coincidence in their trends with respect to \sigmaIRSF\ favors physical interpretations local to the young, dusty star-forming regions which dominate the IR surface density. Indeed, the mean UV interstellar radiation field intensity impinging upon PDRs, $G$ (measured in $\mathrm{G_0=1.6\times10^{-3}\,erg\,s^{-1}\,cm^{-2}}$, \citealt{Habing1968}), relative to the neutral PDR density $\mathrm{n_H}$ increases by 1 dex over \logSigmaIRSF$\,=9-11$ in GOALS as both \cii/\lir\ and \lsix/\lirsf\ fall \citep{DiazSantos2017}.

The low \lpah/\lirsf\ ratios at high \sigmaIRSF\ might indicate a change in the ISM conditions regulating the excitation of both FIR lines and PAH emission within dusty and young star-forming regions \citep[e.g.,][]{DiazSantos2017}. The high $\mathrm{G/n_H}$ ratios found amongst GOALS at \logSigmaIR$\,>10.7$ indicate that the average star-forming region sees a stronger radiation field, which can modify the PAH photoelectric heating efficiency \citep{Bakes1994,Galliano2008,Tielens2008} and lower the radiative coupling between stars and gas \citep{McKinney2020,McKinney2021a}. Changes in the relative heating and cooling could lead to systematically high star-formation efficiency if PAHs photoelectrically convert a lower fraction of energy from the stellar radiation field into gas temperatures \citep{Hollenbach1999,McKinney2021a}. Indeed, simulations that include variable photo-heating laws find 
high photoelectric heating rates suppress star-formation due to excess heating \citep{Forbes2016,Inoguchi2020,Osman2020}. Systematic changes in the photoelectric heating efficiency might leave imprints on the mid-IR spectra as the ionization and/or grain size distribution of PAHs is modified \citep{Draine2001,Maragkoudakis2020}. The overlap between GOALS and $z\sim2$ (U)LIRGs along canonical diagnostic plots of PAH grain properties suggests that the physical mechanisms observed at $z\sim0$ are likely in place and playing an important role at higher$-z$ (Figure \ref{fig:pahs_mdust}, see also \citealt{McKinney2020}). 

\subsection{Dust-obscured galaxies form stars more efficiently at high IR surface densities}
In Figure \ref{fig:discussion} (\textit{Left}) we show the $\mathrm{L_{IR,SF}/M_{dust}}$ ratio vs. \sigmaIRSF. The $\mathrm{L_{IR,SF}/M_{dust}}$ ratio is an empirical tracer of the global star-formation efficiency, which may systematically evolve with redshift \citep{Scoville2017}. As stated, \cite{Kirkpatrick2017} demonstrated that, for fixed distance above the main-sequence, the star-formation efficiency of (U)LIRGs fall along the same relation between $z\sim0-2$. We find a similarly redshift-independent result, 
where high IR luminosity surface density galaxies exhibit more efficient star-formation. Combined with KINGFISH which probes galaxies at lower star-formation rates and more extended sizes, the correlation persists linearly over $\sim5$ orders of magnitude in \sigmaIRSF\ with comparatively shallow increase in star-formation efficiency by one order of magnitude. \cite{Kirkpatrick2017} find no correlation between $\mathrm{L_{IR,SF}/M_{dust}}$ and the ISM extent measured from CO, radio, or Pa$\alpha$ for a handful of high$-z$ galaxies and local LIRGs in GOALS \citep{Rujopakarn2011}. As shown in Figure \ref{fig:sfe_reff} and Table \ref{tab:par}, we find a statistically significant correlation between $\mathrm{L_{IR,SF}/M_{dust}}$ and the uniformly measured effective radii from far-IR wavelengths in local galaxies, but not at $z\sim2$. 
A much stronger trend manifests at all redshifts using \sigmaIRSF, consistent with multiple studies in the literature demonstrating that luminosity surface densities more accurately reflect the gas and star-formation conditions in galaxies rather than the total luminosity or size alone \citep[e.g.,][]{Rujopakarn2011,DiazSantos2017,Elbaz2018,McKinney2020,DiazSantos2021}. 

The star-formation efficiencies are comparable between GOALS and the $z\sim2$ (U)LIRGs in this study; however, the samples are different in terms of their position relative to main-sequence star-formation for their corresponding epochs \citep{Lutz2016,Kirkpatrick2017}. Using the main-sequence parameterization from \cite{Speagle2014}, we show in Figure \ref{fig:discussion} (\textit{Right}) the ratio of each galaxy's specific star-formation rate (sSFR$\,\equiv\mathrm{SFR/M_*}$) to the main-sequence sSFR for the corresponding stellar mass and redshift against \sigmaIRSF.

GOALS galaxies are typically a factor of $\sim10$ above the main-sequence compared to a factor of $\sim4$ in ALMA-detected $z\sim2$ (U)LIRGs. (U)LIRGs at $z\sim2$ span a range of nearly two orders of magnitude in \sigmaIRSF\ with a similar dispersion in distance from the main-sequence $\sim0.2$ dex as found locally, albeit shifted down. High$-z$ (U)LIRGs can exhibit high and low dust-obscured star-formation rate surface densities for fixed position along the main-sequence, whereas \sigmaIRSF\ in local (U)LIRGs correlates with distance from the main-sequence \citep{Elbaz2011,Lutz2016}. Indeed, \cite{GomezGuijarro2022b} even find high star-formation surface densities in main-sequence, mm-selected galaxies at $z\sim1-3$ that also exhibit high star-formation efficiencies. Such galaxies could remain on the main-sequence for 300 Myr - 1 Gyr \citep{Ciesla2022}. The comparable star-formation efficiencies (Fig.~\ref{fig:discussion} \textit{Left}) and dust properties (Fig.~\ref{fig:sigma_PAH}) show that the star-forming cores of GOALS galaxies are good local analogs to high$-z$ (U)LIRGs in terms of their general star-formation properties despite both populations occupying a fundamentally different region with respect to the main-sequence. 

The relation between $\mathrm{L_{IR,SF}/M_{dust}}$ and \sigmaIRSF\ is consistent with the \sigmaIRSF\ dependence of both \lpah/$\mathrm{M_{dust}}$ (Fig.~\ref{fig:pahs_mdust}) and \lpah/\lirsf\ (Fig.~\ref{fig:sigma_PAH}). While these scaling relations illustrate some link between the total dust mass, PAH emission, and star-formation surface densities, the underlying physical association is not yet clear. This is especially true for the complex role played by dust grains in the ISM. Constant \lpah/$\mathrm{M_{dust}}$ suggests that relative to the total dust mass, PAHs are not systematically destroyed in (U)LIRGs at high IR surface densities by strong radiation fields. 
The low \lpah/\lirsf\ ratios at high \sigmaIRSF\ could be causally linked to the high star-formation efficiencies if the PAHs consistently trace the total dust at all IR surface densities (i.e., $\mathrm{L_{PAH}}\propto M_{dust}\Rightarrow \mathrm{L_{IR}/M_{dust}\sim L_{IR}/L_{PAH}\propto SFE}$). Indeed, PAH line ratios tracing the ionization state of grains exhibit little scatter amongst $z=0$ (U)LIRGs \citep[e.g.,][]{Stierwalt2014} which could otherwise change the PAH mass-to-light ratio; however, the scatter in the same ratios is higher at $z\sim2$ \citep{McKinney2020} and some change in the size distribution of grains is apparent when accounting for the 3.3$\,\mu$m PAH feature at $z=0$ \citep{McKinney2021a}. \textit{JWST}/MIRI MRS could test the PAHs in further detail through high SNR mid-IR spectroscopy. A complimentary approach would be to follow-up \textit{Spitzer}/IRS targets with ALMA to measure far-infrared lines like \cii\ and investigate the potential link between low \cii/PAH ratios and dust grain properties. This would clarify whether or not changes to the grain properties (i.e., size, charge, PAH mass fraction) are important in regulating the formation of stars within distant and dusty galaxies.

\section{Conclusion\label{sec:conclusions}}
Using the ALMA archive, we measure IR sizes (\Reff) and dust masses for a sample of 15 $z\sim1-3$ (U)LIRGs in GOODS-S with existing \textit{Spitzer}/IRS spectroscopy. We compare these high-redshift galaxies to local luminous IR galaxies in the GOALS survey \citep{Armus2009} and KINGFISH galaxies with more typical star-formation rates \citep{Kennicutt2011}. We combine the size and dust mass measurements with mid-IR spectral features to assess the degree by which changes in IR surface density drive changes in the total ISM content and conditions. Our main conclusions are as follows:

\begin{enumerate}
    \item The total dust mass scales with \Reff\ amongst $z\sim0-3$ (U)LIRGs along an average dust mass surface density of $\Sigma_{\mathrm{dust}}\sim10\,\mathrm{M_\odot\,pc^{-2}}$, corresponding to an average gas mass surface density of $\Sigma_{\mathrm{gas}}\sim1000\,\mathrm{M_\odot\,pc^{-2}}$ assuming a gas-to-dust ratio of 100. We find IR sizes that are smaller for fixed dust mass relative to typical star-forming galaxies selected from the simulated galaxies in TNG50, but the $2\sigma$ simulation outliers do reproduce this parameter space of high IR surface density dusty galaxies. 
    \item (U)LIRGs at $z\sim1-3$ with measurable AGN contribution to the mid-IR exhibit preferentially smaller far-IR sizes and larger IR surface densities.
    \item The $\mathrm{L_{PAH}}$/$\mathrm{M_{dust}}$ ratio amongst (U)LIRGs does not evolve with redshift suggesting similar PAH mass fractions today and at cosmic noon. PAHs should be used with caution as a total dust mass tracers for (U)LIRGs given the $\pm0.3$ dex scatter about the mean in $\mathrm{L_{PAH}}$/$\mathrm{M_{dust}}$; however, this scatter is significantly lower for KINGFISH galaxies at \logSigmaIRSF$\,<9.5$ where $\mathrm{L_{PAH}}$/$\mathrm{M_{dust}}$ correlates with \sigmaIRSF.
    \item We find an anti-correlation between the $\mathrm{L_{PAH}}$/\lirsf\ ratio and \sigmaIRSF\ at $z\sim2$ reminiscent of FIR cooling line deficits and down to lower IR surface densities than previously probed. This suggests that changes in ISM conditions regulating the PAH and FIR line emission for $z\sim0$ galaxies are also likely present in the $z\sim2$ LIRG population. 
    \item (U)LIRGs with higher IR surface densities show larger \lirsf/$\mathrm{M_{dust}}$ ratios, emblematic of more efficient star-formation. We find a trend between \sigmaIRSF\ and the star-formation efficiency spanning five orders of magnitude in \sigmaIRSF\ for $z\sim0$ and $z\sim2$ galaxies. This trend extends linearly from the KINGFISH galaxies which have lower \lir\ and more extended IR sizes than the LIRGs and ULIRGs.
    \item Whereas $z\sim0$ galaxies at high IR surface density tend to be further above the main-sequence, we find a weaker correlation between \sigmaIRSF\ and the distance above the star-forming main-sequence at $z\sim2$. In other words, the extent of star-formation for fixed distance above the main-sequence is more varied at $z\sim2$ than what is found locally. This is consistent with previous works finding high star-formation surface densities for $z\sim2$ dusty galaxies on the star-forming main-sequence. 
\end{enumerate}

Despite occupying fundamentally different regimes with respect to the star-forming main-sequence, the ratios between \lirsf, total dust mass, PAHs and IR sizes in high- and low-redshift (U)LIRGs are similar. Taken together, the results of this analysis paint a picture of dust-obscured galaxy formation in which the gas and dust conditions set the star-formation conditions, which can increase the star-formation efficiency and support large star-formation rate surface densities. Dust-obscured star-formation at earlier cosmic times may be a scaled up version of what is found locally in the star-forming cores of IR-luminous galaxies, albeit sustained over larger areas and fed by increased gas fractions.

\begin{acknowledgments}
\footnotesize
We thank the referee for their insightful comments that strengthened the quality of the paper.
We thanks the ASAGAO and GOODS-ALMA teams for their efforts to make these surveys possible.
J.M. is supported by the ALMA archival SOS grant (SOSPADA-011), which made this research possible. HI acknowledges support from JSPS KAKENHI Grant Number JP19K23462 and JP21H01129. This work is based on observations made with the Herschel Space Observatory, a European Space Agency (ESA) Cornerstone Mission with science instruments provided by European-led Principal Investigator consortia and significant participation from NASA. The Spitzer Space Telescope is operated by the Jet Propulsion Laboratory, California Institute of Technology, under NASA contract 1407. This research has made use of the NASA/IPAC Extragalactic Database (NED), which is operated by the Jet Propulsion Laboratory, California Institute of Technology, under contract with the National Aeronautics and Space Administration, and of NASAs Astrophysics Data System (ADS) abstract service. This research has made use of the NASA/IPAC Infrared Science Archive, which is operated by the Jet Propulsion Laboratory, California Institute of Technology, under contract with the National Aeronautics Space Administration.
\end{acknowledgments}

\appendix
\section{Image Cutouts \label{app:A}}

\begin{figure*}
    \centering
    \gridline{\fig{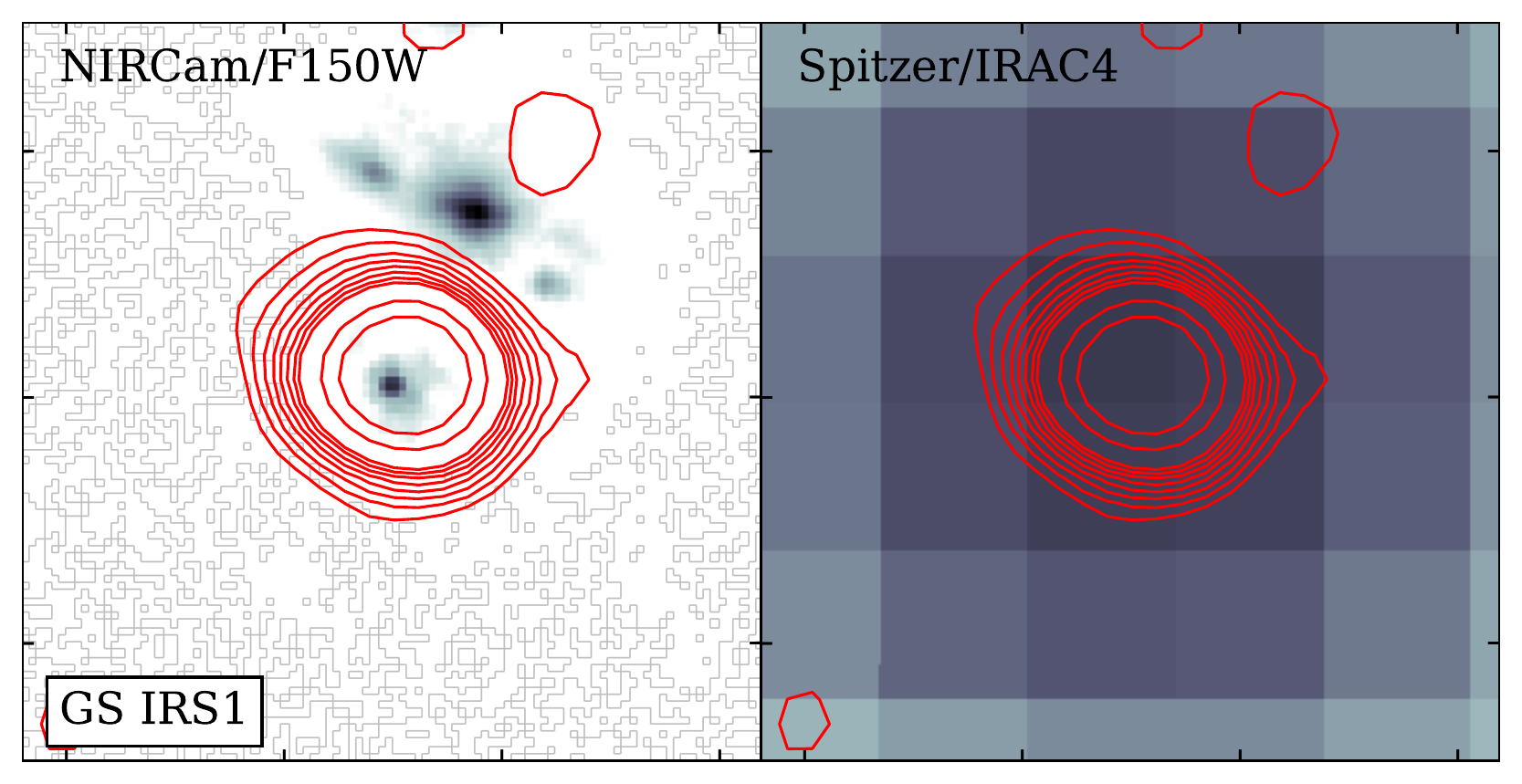}{0.49\textwidth}{}
              \fig{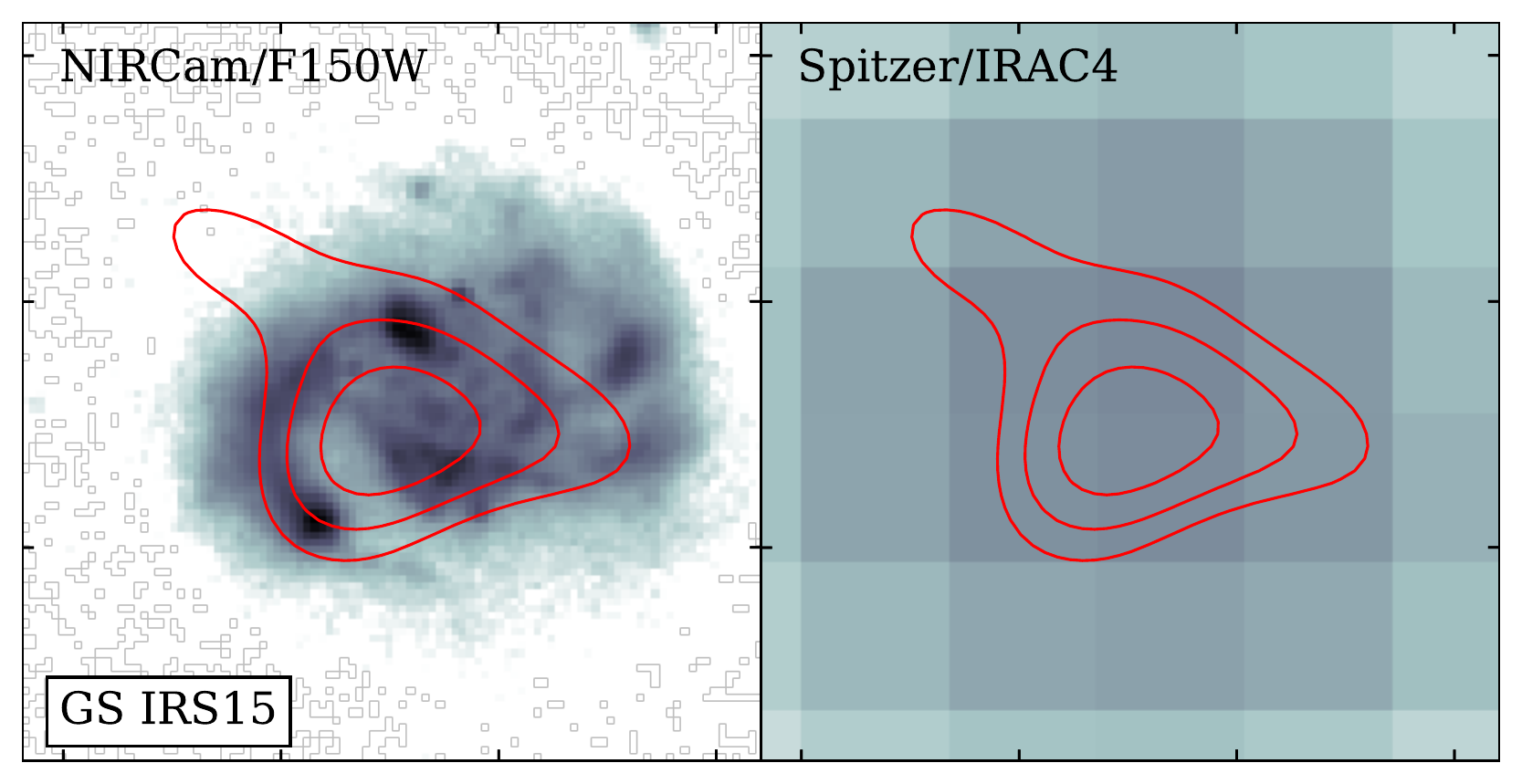}{0.49\textwidth}{}
              }\vspace{-30pt}
    \gridline{\fig{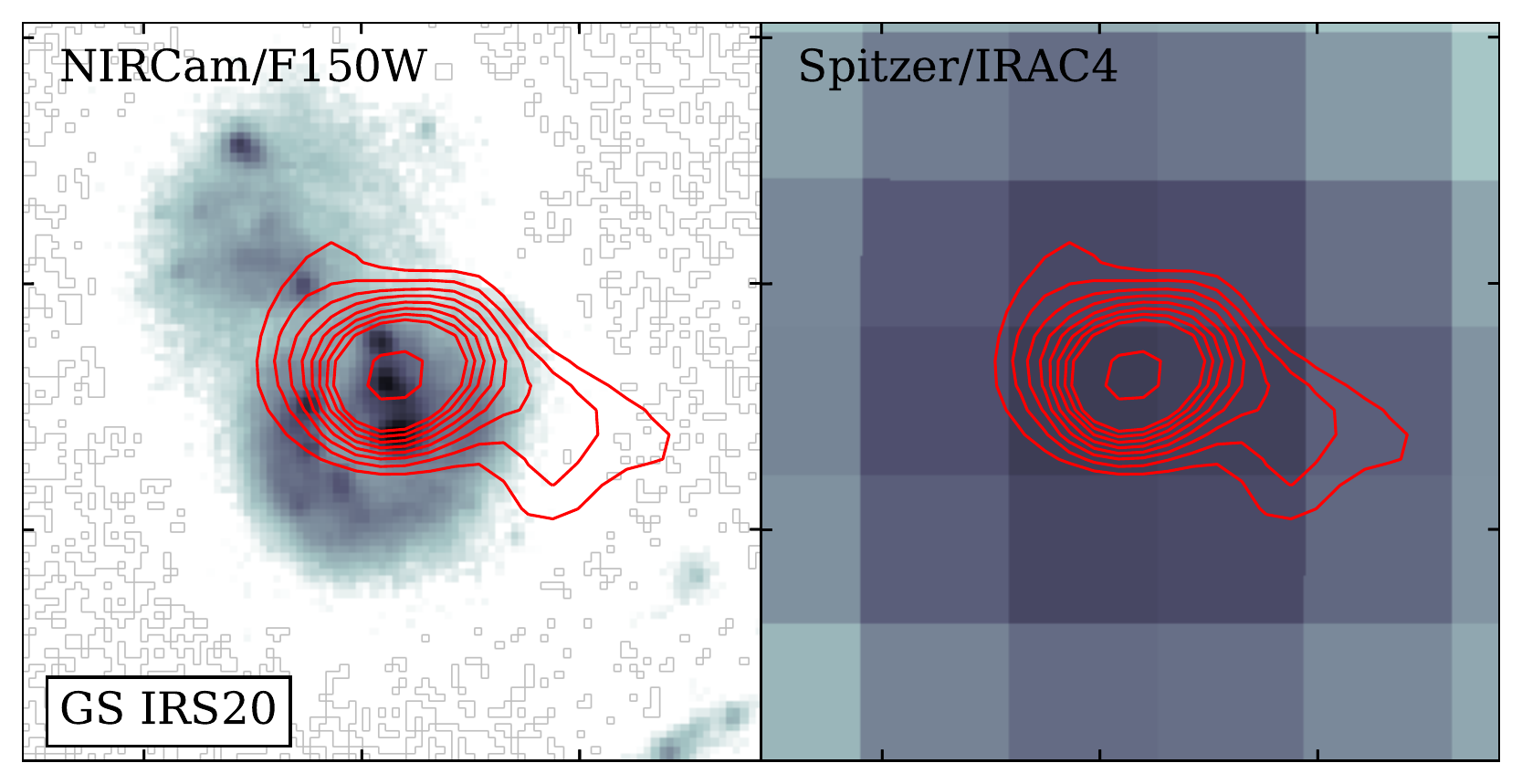}{0.49\textwidth}{}
              \fig{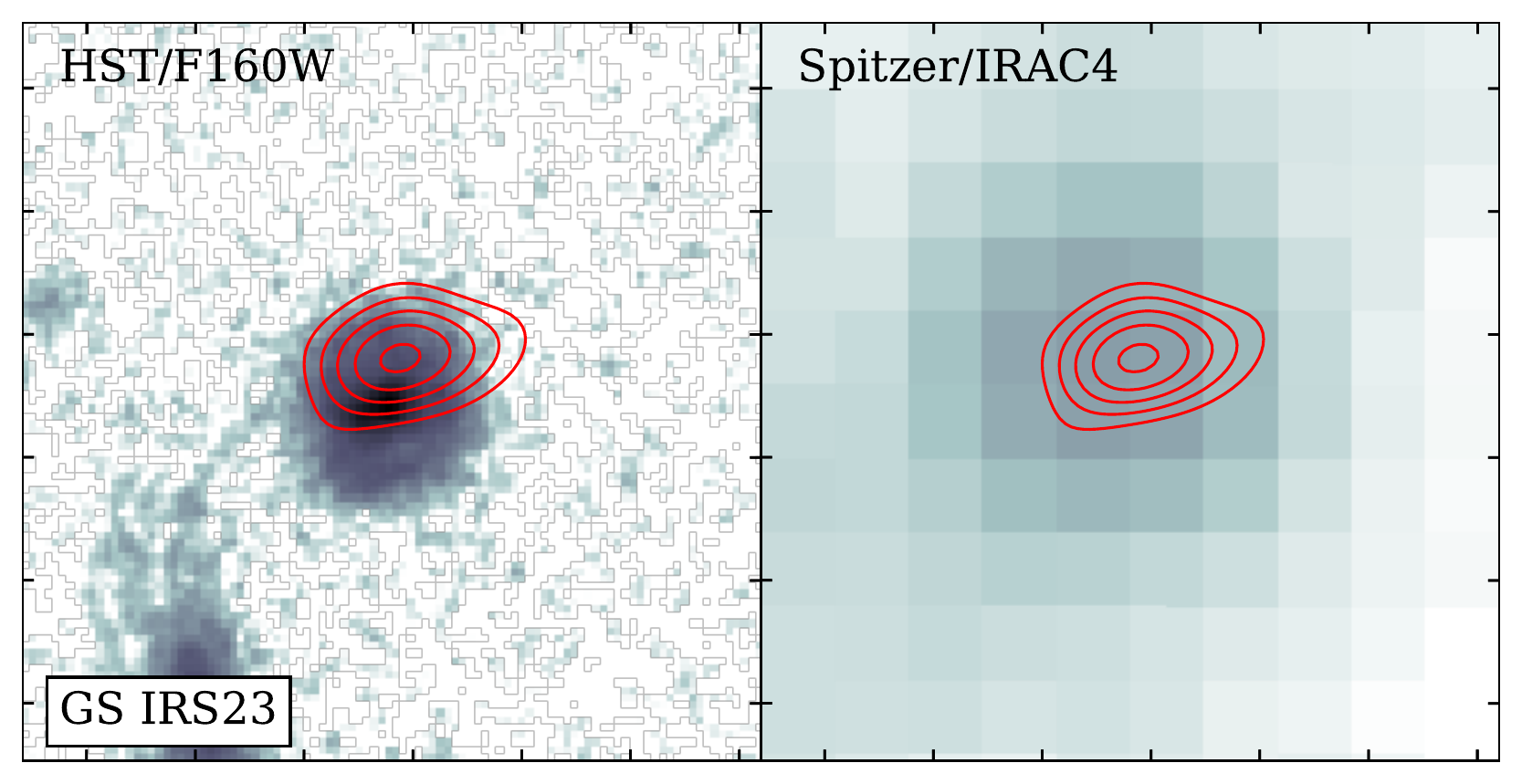}{0.49\textwidth}{}
              }\vspace{-30pt}
    \gridline{\fig{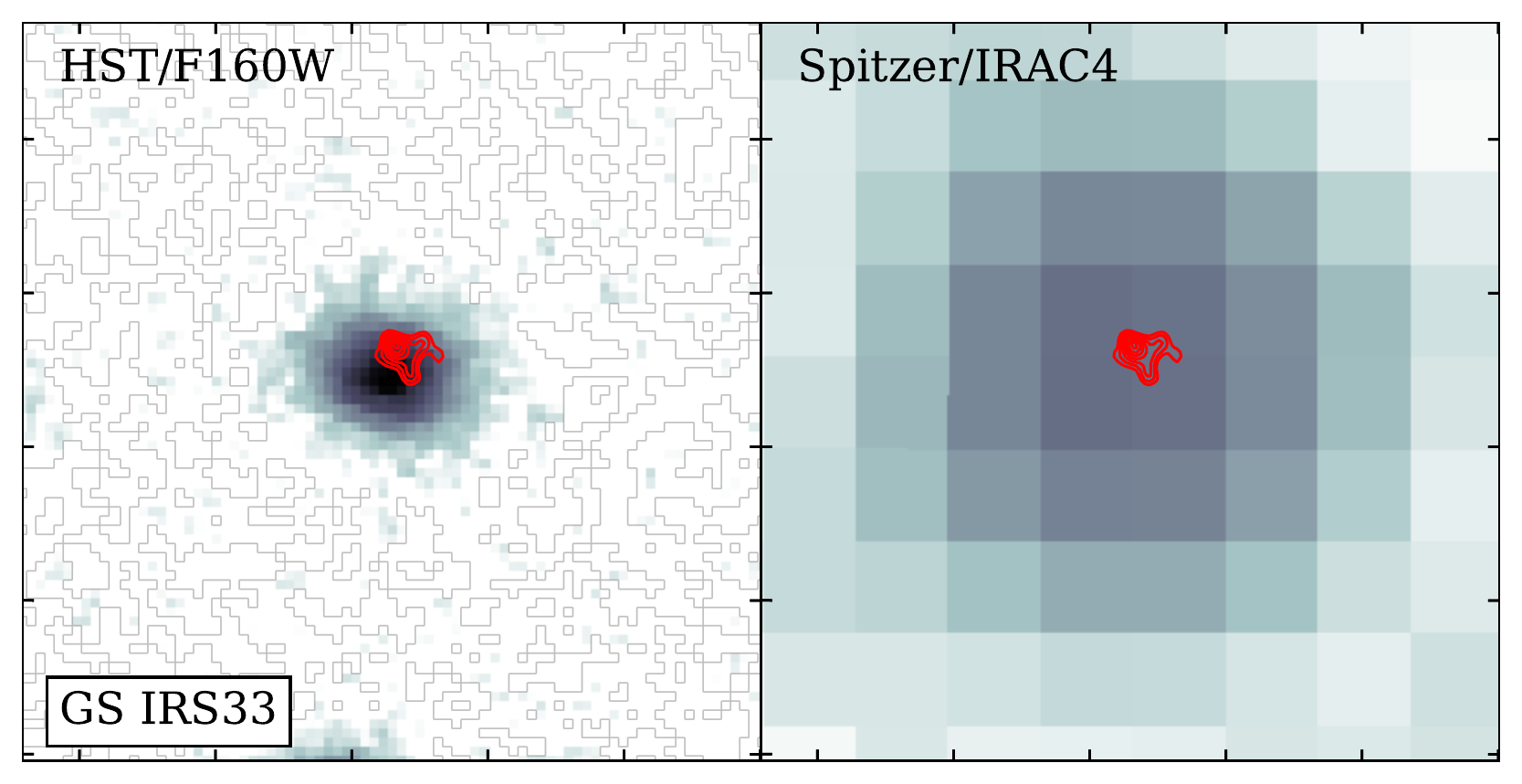}{0.49\textwidth}{}
              \fig{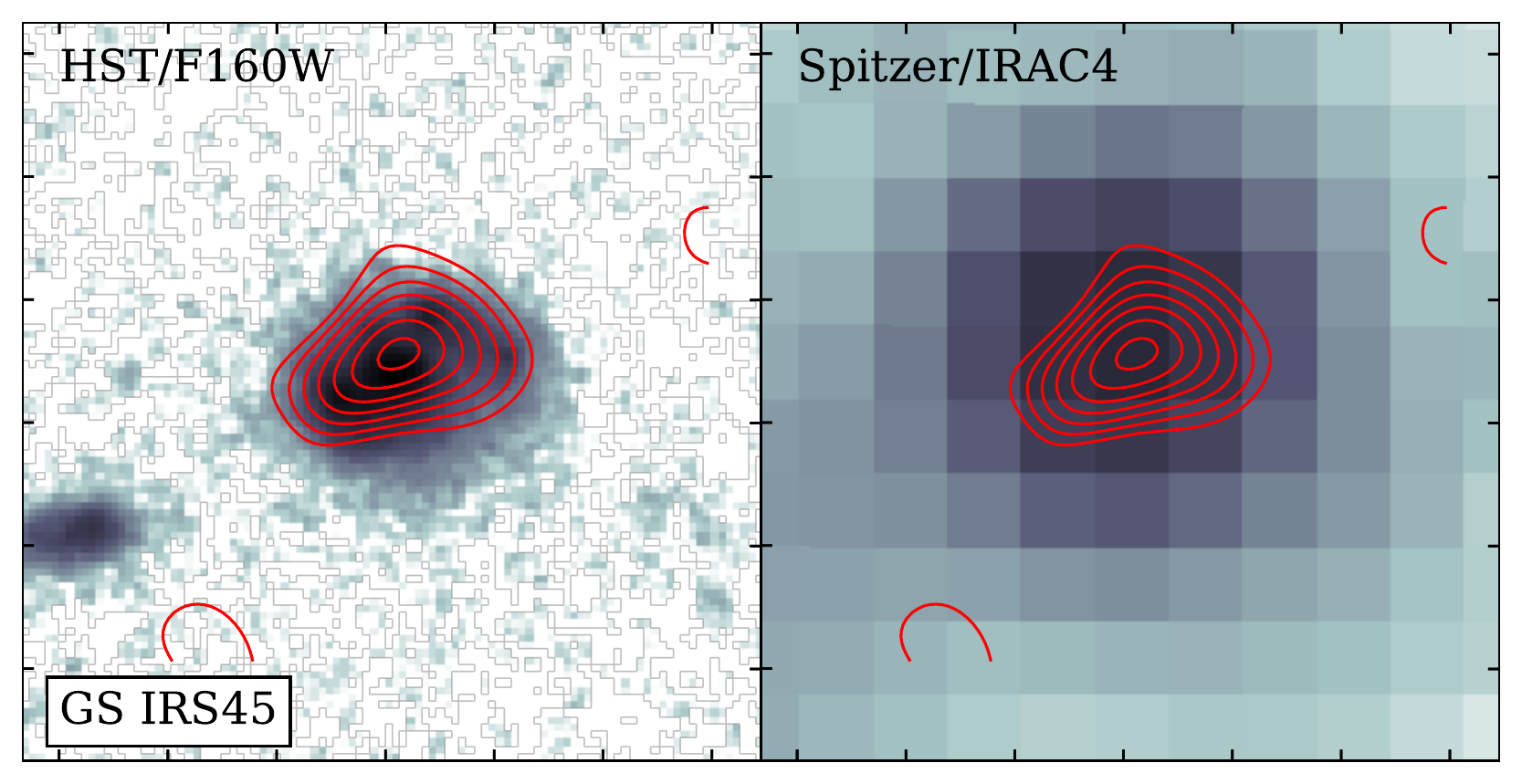}{0.49\textwidth}{}
              }\vspace{-30pt} 
    \gridline{\fig{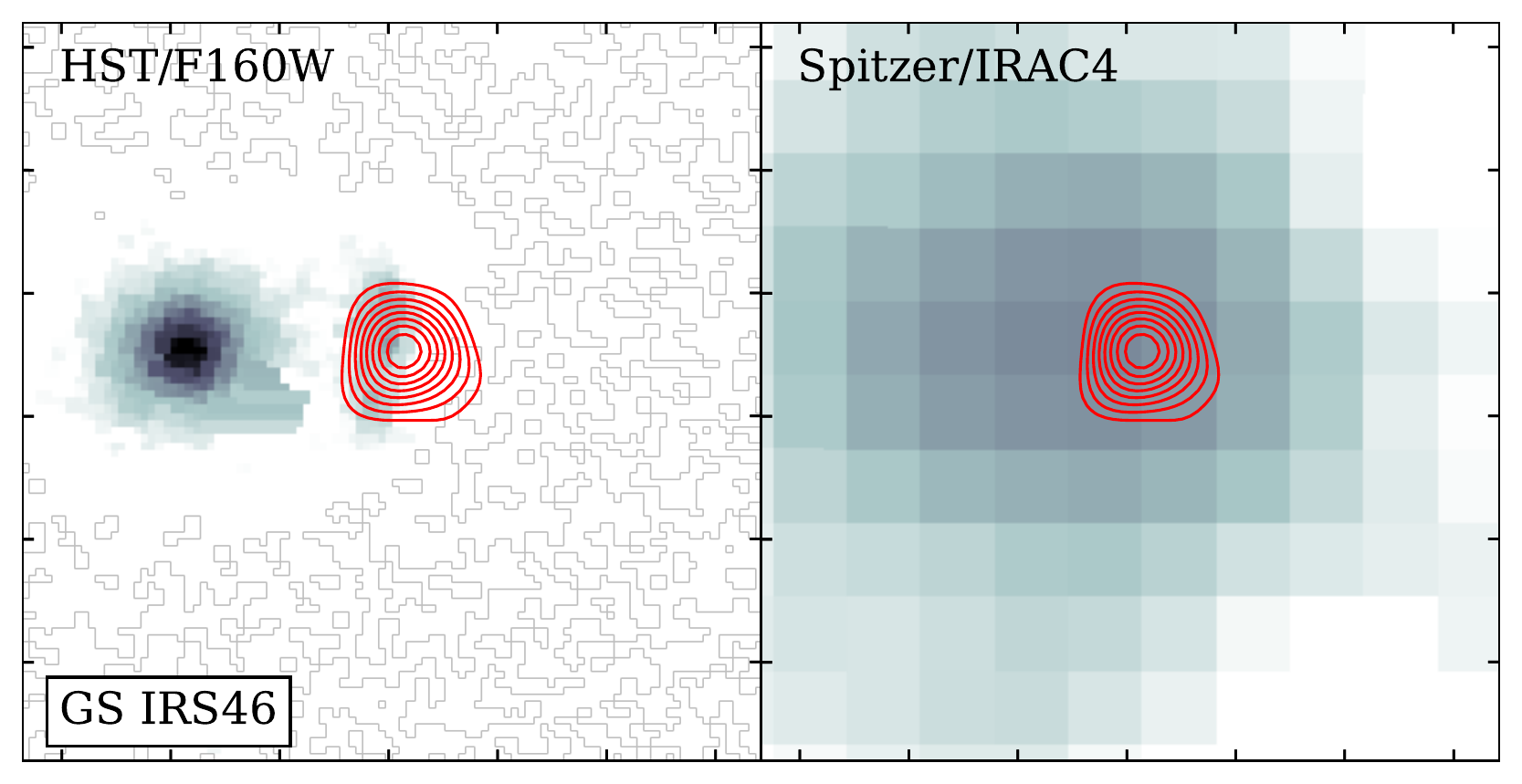}{0.49\textwidth}{}
              }\vspace{-5pt}
    \caption{ALMA contours (red levels) over \textit{JWST}/NIRCam F150W imaging from the JADES survey \citep{Eisenstein2023,Rieke2023,JADES} where available, and \textit{HST}/F160W otherwise. We also show the ALMA contours over \textit{Spitzer}/IRAC4 images. The ALMA contours are drawn at $3\sigma, 4\sigma, 5\sigma, 10\sigma, 15\sigma, 20\sigma$. Tick marks are spaced on each cutout by $1.^{\prime\prime}0$.}
    \label{fig:cutouts}
\end{figure*}

\begin{figure*}
    \centering
    \gridline{\fig{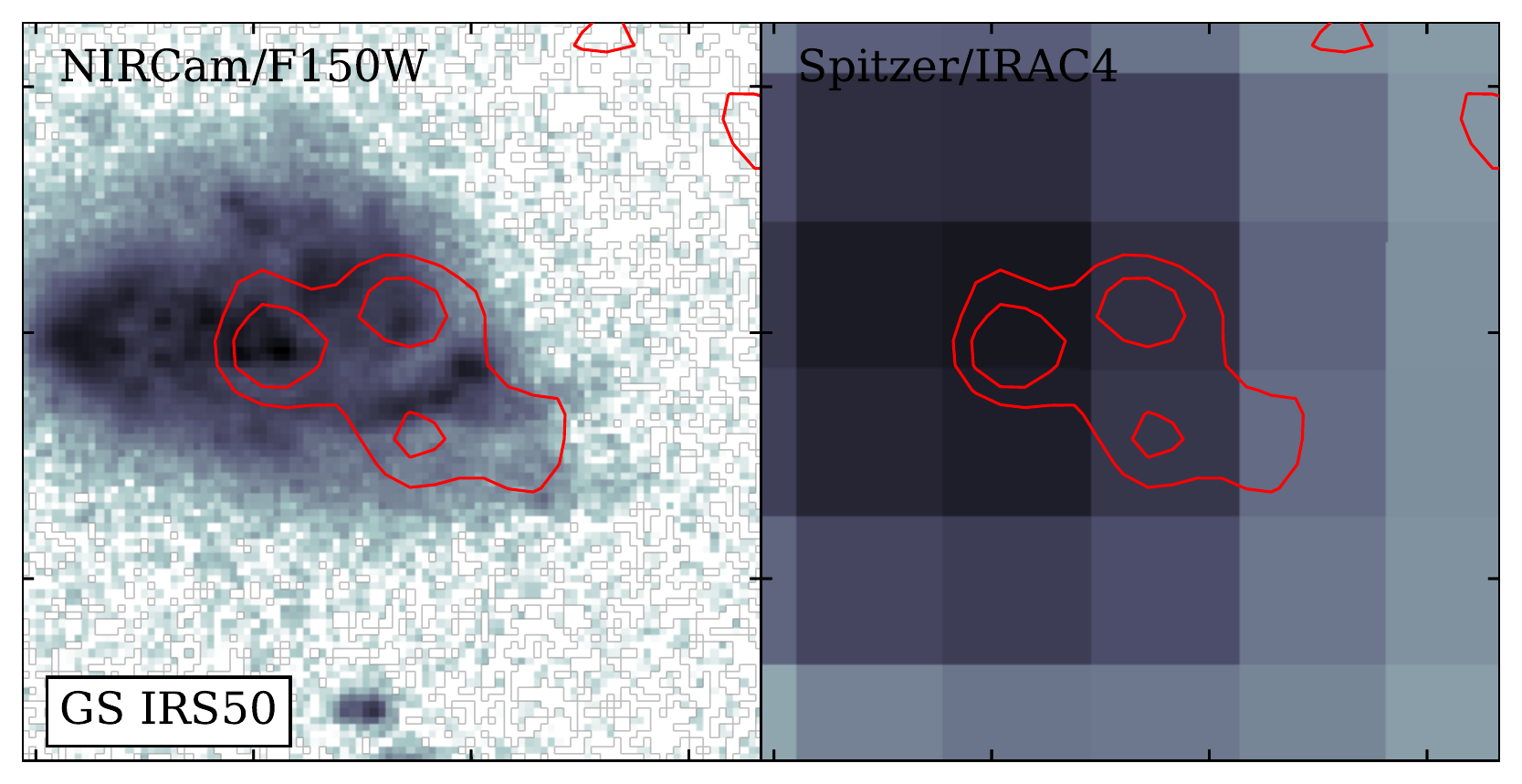}{0.49\textwidth}{}
              \fig{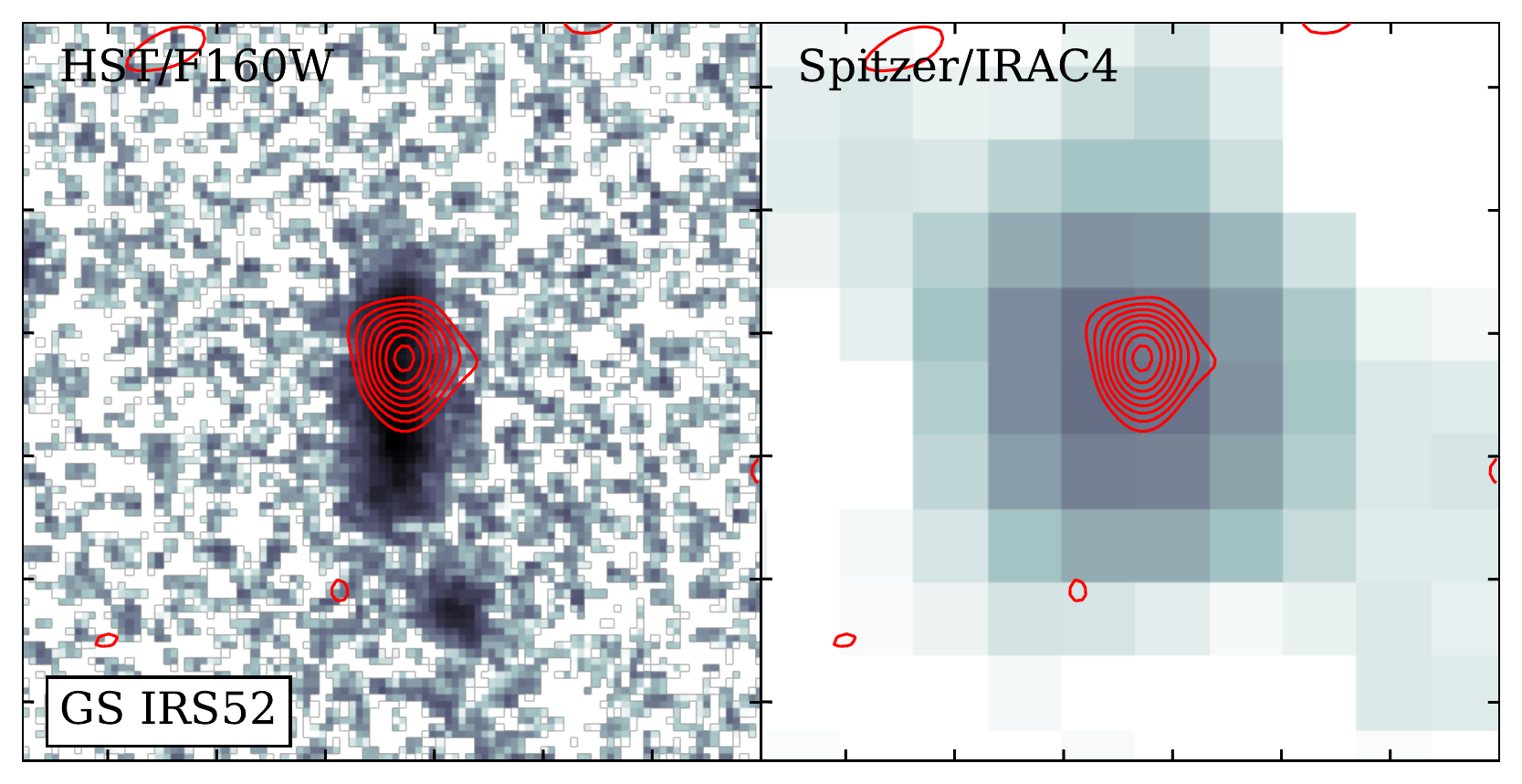}{0.49\textwidth}{}
              }\vspace{-30pt}
    \gridline{\fig{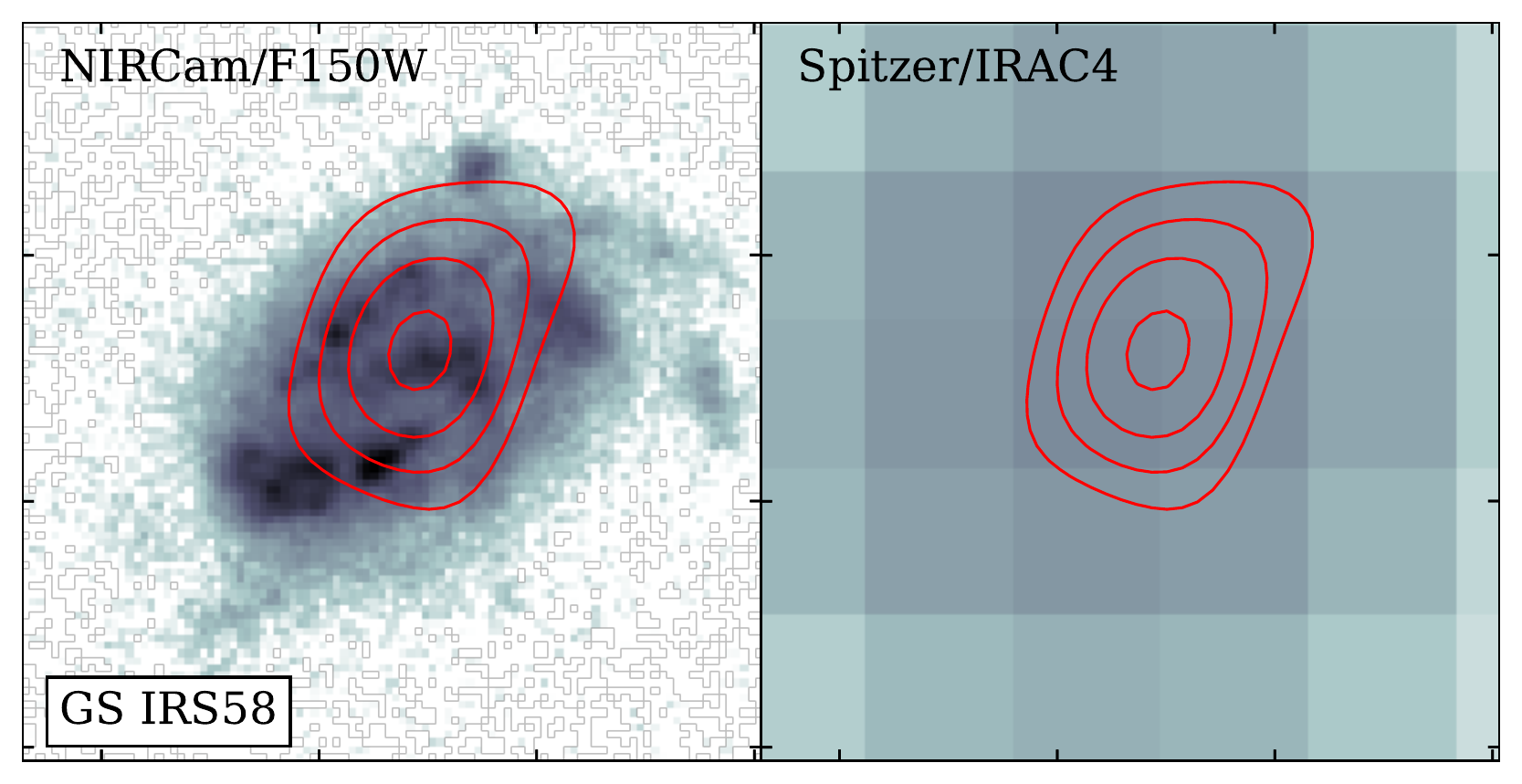}{0.49\textwidth}{}
              \fig{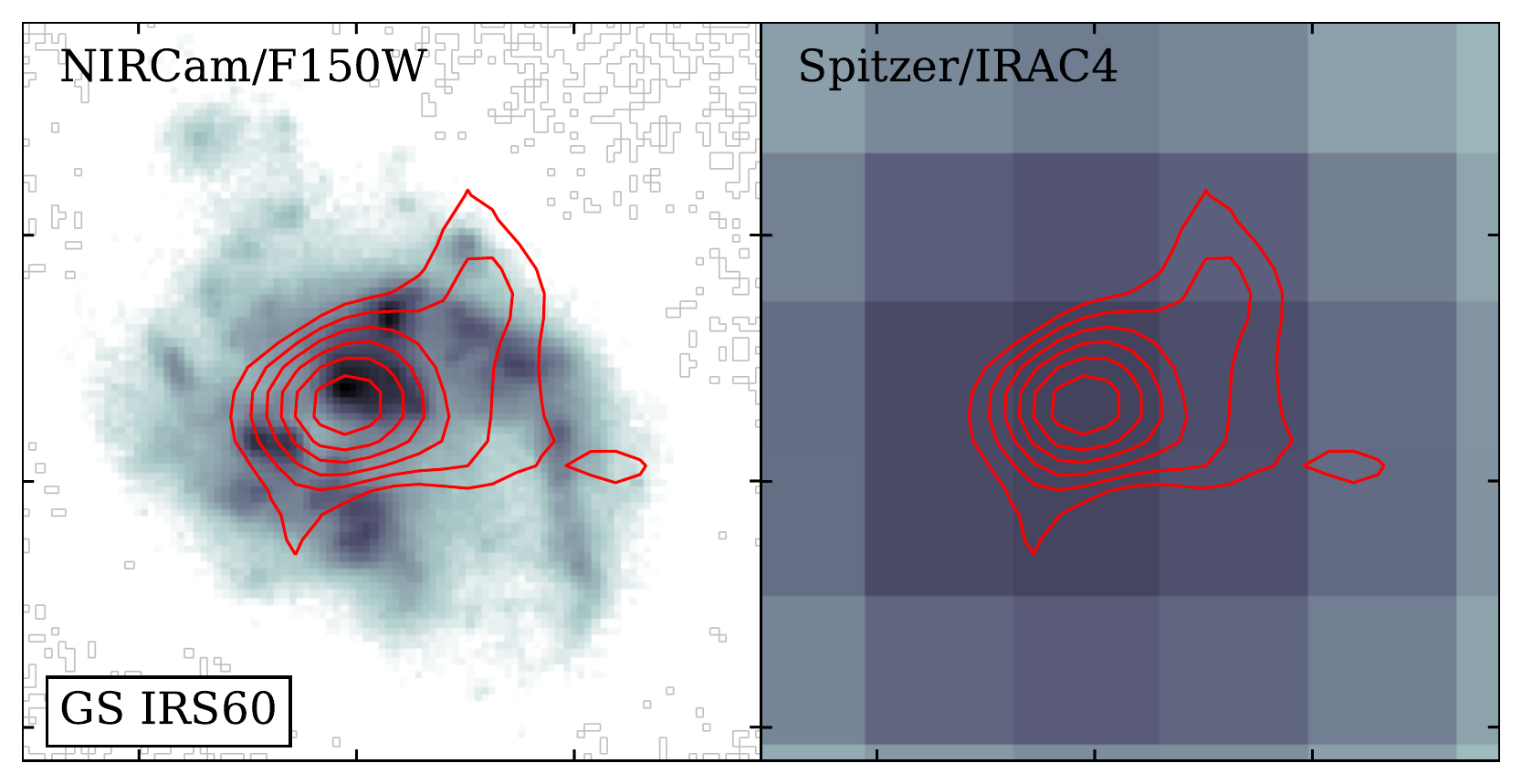}{0.49\textwidth}{}
              }\vspace{-30pt}
    \gridline{\fig{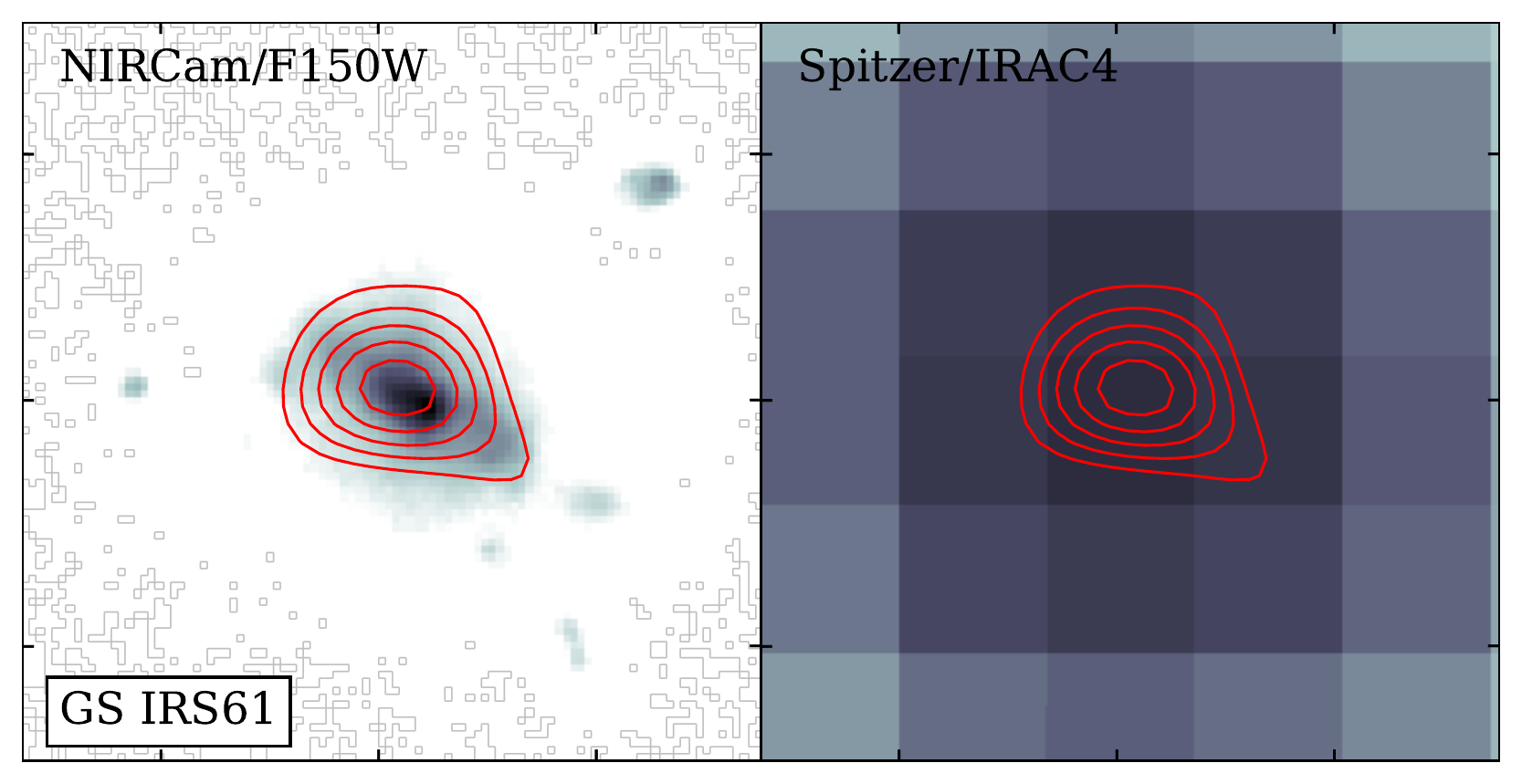}{0.49\textwidth}{}
              \fig{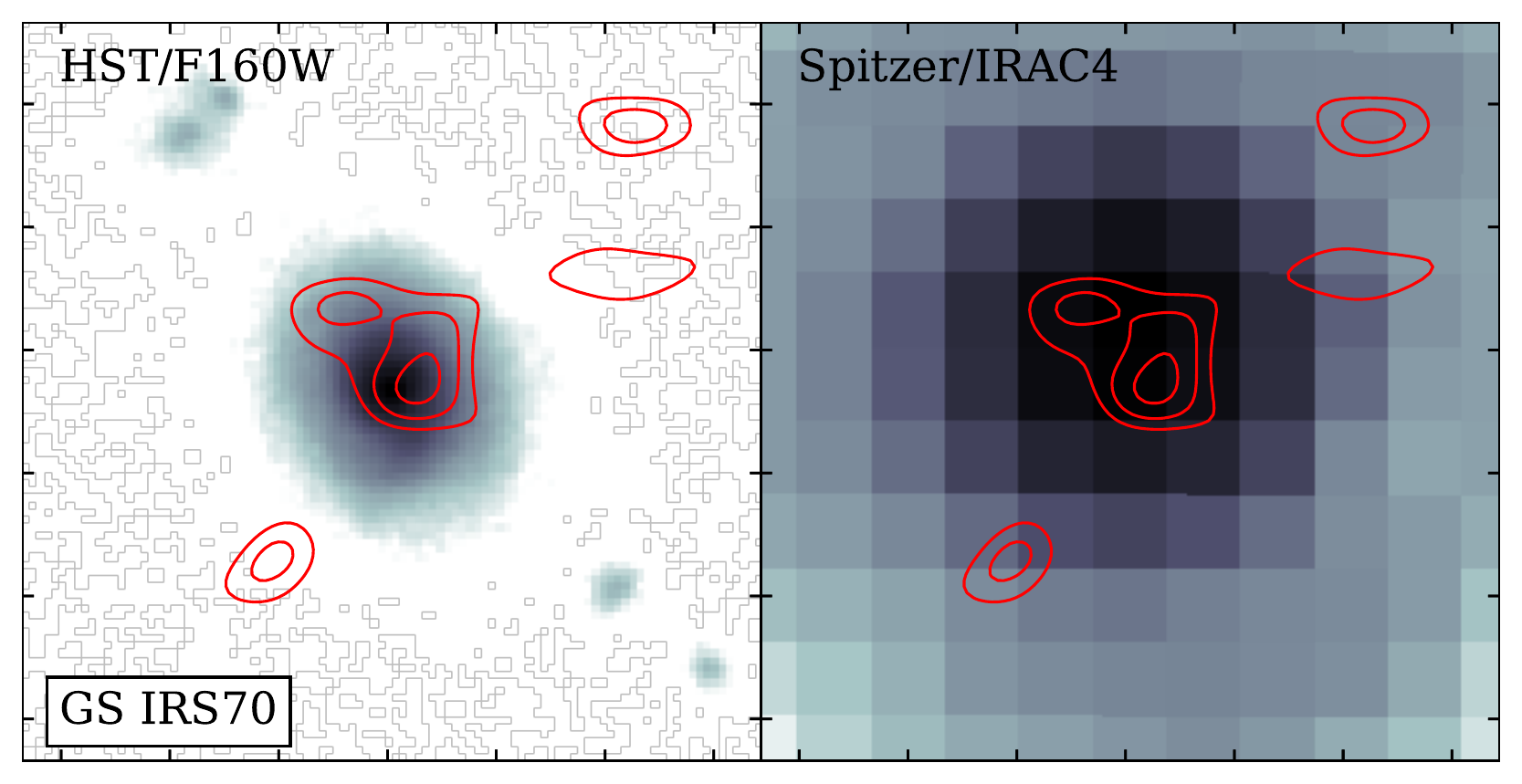}{0.49\textwidth}{}
              }\vspace{-30pt} 
    \gridline{\fig{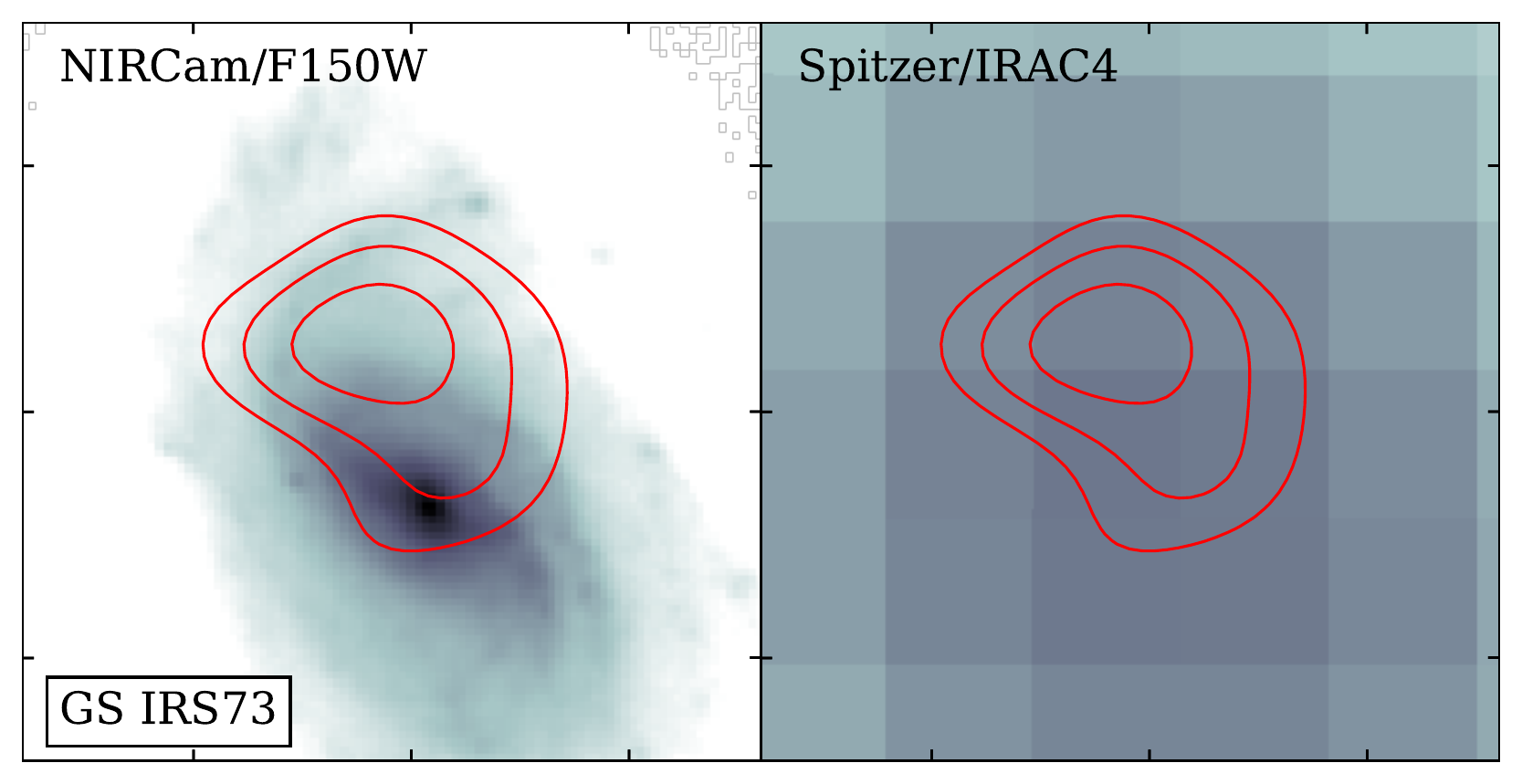}{0.49\textwidth}{}
              \fig{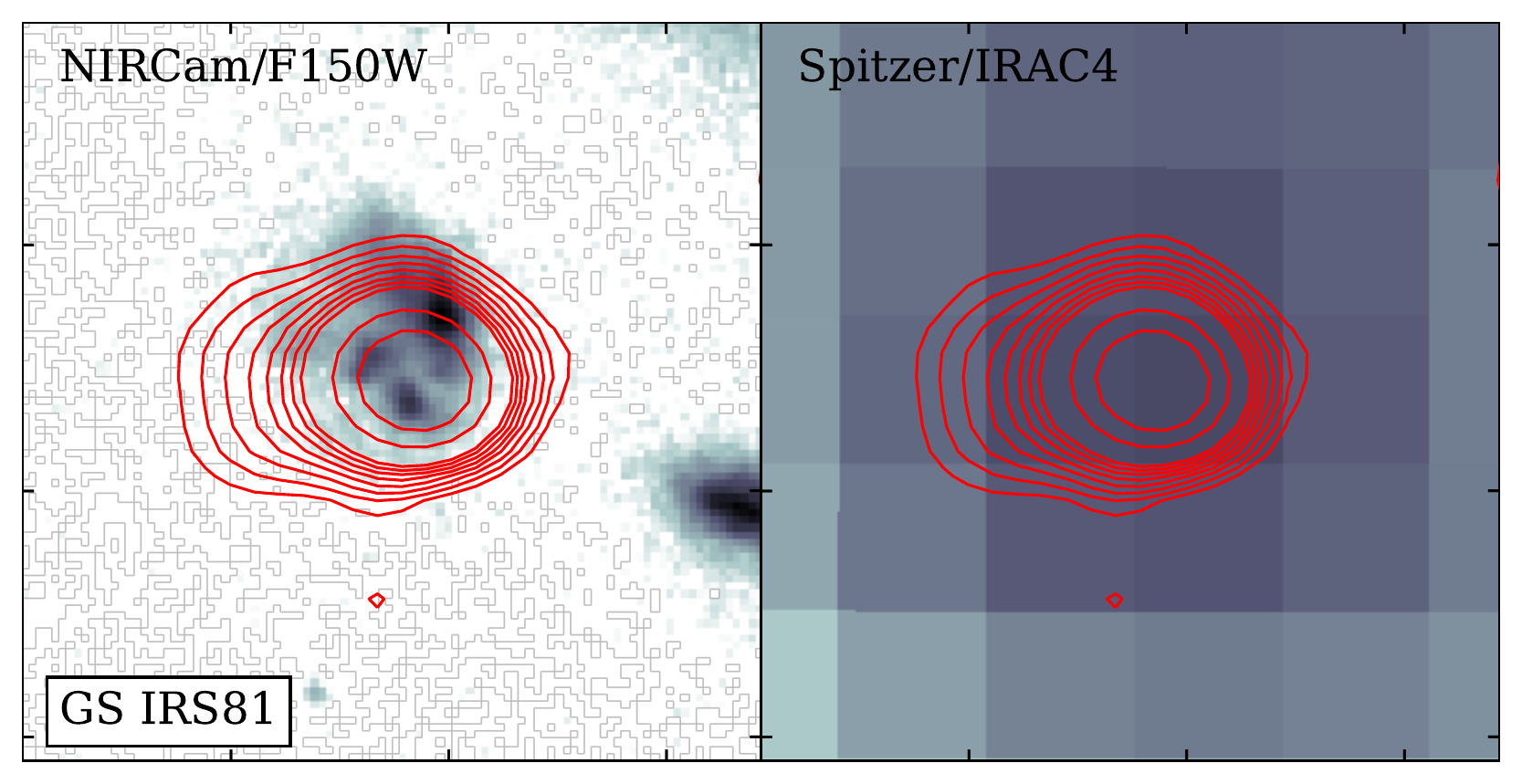}{0.49\textwidth}{}
              }\vspace{-10pt}
       \caption{Continuation of Fig.~\ref{fig:cutouts}.}
    \label{fig:cutouts2}
\end{figure*}

\clearpage
\bibliography{references.bib}

\end{document}